Chapter 33

# Laser Processing for Energy Applications

**Marcos Soldera[1], Robert Baumann[1], Bogdan Voisiat[1], Wilhelm Pfleging[2,*], Penghui Zhu[2], Jörn Bonse[3,*], Andrés Fabián Lasagni[1,4,*]**

[1] Institute of Manufacturing Technology, Technische Universität Dresden, George-Bähr Str. 3c, 01069 Dresden, Germany

[2] Institute of Applied Materials – Applied Physics (IAM-AWP), Karlsruhe Institute of Technology (KIT), Hermann-von-Helmholtz-Platz 1, 76344 Eggenstein-Leopoldshafen, Germany

[3] Bundesanstalt für Materialforschung und -prüfung (BAM), Unter den Eichen 87, 12205 Berlin, Germany

[4] Fraunhofer Institute for Material and Beam Technology IWS, Winterbergstr. 28, 01277 Dresden, Germany

e-mail: marcos.soldera@tu-dresden.de; robert.baumann1@tu-dresden.de; bogdan.voisiat@tu-dresden.de; wilhelm.pfleging@kit.edu; penghui.zhu@kit.edu; joern.bonse@bam.de; andres_fabian.lasagni@tu-dresden.de

*corresponding authors

## Abstract

Laser processing has emerged as a transformative technology in the energy sector, enabling precise, efficient, and scalable solutions for energy generation, storage, and sustainability. This chapter explores the diverse applications of laser technology in energy applications. It begins with an overview of laser-driven inertial confinement fusion for energy production, highlighting key experimental facilities and crucial laser parameters. The role of lasers in solar energy harvesting is examined, including photovoltaic cell optimization, concentrated solar power, and laser-assisted recycling of solar modules. The chapter further reviews recent developments into laser-based techniques for enhancing electrochemical performance through surface texturing at various scales. Advancements in lithium-ion battery technology are also discussed, focusing on next-generation materials, battery manufacturing processes, and the impact of laser techniques on performance, safety, and scalability. The chapter concludes with insights into future prospects for laser applications in sustainable energy solutions.

# Contents

# 1. Introduction

As the global demand for sustainable and efficient energy solutions grows, laser technology has emerged as an important tool for advancing energy generation, storage, and conversion. These areas are explored in this chapter, where laser processing plays a transformative role with respect to the treated materials state, shape, and structure.

One of the most ambitious applications is laser fusion for energy generation, where high-energy lasers drive inertial confinement fusion (ICF) as a potential pathway to achieving sustainable nuclear fusion. In this frame, Sect. 2 focuses into the fundamental principles of fusion energy, from historical discoveries to the latest advancements in laser-driven fusion experiments.

Beyond fusion, laser technology in solar energy harvesting has shown to play a crucial role for instance in enhancing the efficiency of photovoltaic cells and concentrated solar power systems. This field is discussed in Sect. 3. Laser-based processes, such as laser scribing and texturing, are used to optimize light absorption and improve solar module performance. Additionally, lasers reduce the material consumption and facilitate the recycling of solar panels, addressing sustainability concerns in renewable energy technologies. These advancements contribute to making solar energy more efficient, cost-effective, and environmentally friendly.

Another critical area of application is laser techniques for improved electrochemical performance (Sect. 4), where laser surface structuring enhances electrode properties for energy storage and conversion. Techniques such as Direct Laser Writing (DLW), Direct Laser Interference Patterning (DLIP), and Laser-Induced Periodic Surface Structures (LIPSS) modify material surfaces to improve electrochemical reactions and lower energy losses. These methods have shown significant promise in applications like hydrogen evolution reactions and other energy carrier technologies, providing innovative solutions to enhance energy efficiency.

Finally, laser technology for advanced battery systems will be crucial in the development of next-generation lithium-ion batteries, as explored in Sect. 5. Lasers contribute to various stages of battery manufacturing, from electrode structuring to material processing, improving battery performance, safety, and longevity. Ultrafast laser techniques enable precise modifications at the micro- and nanoscale, enhancing charge transport and energy density while facilitating large-scale production. All these topics are described in this chapter.

# 2. Laser Fusion for Energy Generation

Fusion represents the preeminent reaction within our observable universe, serving as the fundamental process that fuels both our Sun and other stellar bodies. Nonetheless, the theoretical frameworks and principles underlying nuclear fusion began to crystallize only in the 1920s, when British astrophysicist Arthur Eddington postulated that stars obtain their energy through the fusion of hydrogen into helium [Eddington, 1920]. In the aftermath of Eddington's seminal publication, Robert d'Escourt Atkinson and Fritz Houtermans conducted the initial quantitative assessments regarding the rate of nuclear fusion occurring in stars [Atkinson, 1929]. Concurrently, Ernest Rutherford was investigating the atomic structure. Through his renowned experiment conducted in 1934, Rutherford demonstrated the fusion of deuterium into helium, observing that "an enormous effect was produced" during this transformative process.

His protégé, Mark Oliphant, employed an enhanced iteration of the experimental apparatus, utilizing deuterium instead of hydrogen, and subsequently identified helium-3 and tritium, thus providing evidence that heavy hydrogen nuclei could engage in nuclear reactions. This represented the inaugural direct demonstration of fusion within a laboratory setting [Oliphant, 1934].

Fusion reactions are governed by the interaction between two fundamental forces: the electrostatic force, which causes atomic nuclei to repel each other, and the strong nuclear force, which binds them together when they come close enough. For nuclei lighter than iron-56, such as hydrogen, fusion is exothermic, meaning it releases energy. Hydrogen, having a single proton, is the easiest element to fuse and offers the highest net energy output. This makes it an ideal candidate for fusion fuel.

To initiate fusion, fuel atoms must be provided with enough kinetic energy to overcome the "Coulomb barrier," the energy threshold required to bring nuclei close enough for the strong force to take effect. This can be achieved by heating atoms to extremely high temperatures, causing them to lose their electrons and form a hot plasma, an electrically conductive cloud of ions and free electrons. The unique properties of plasma allow it to be manipulated and confined using magnetic fields, a principle employed by various fusion devices [McCracken, 2005].

The pursuit of nuclear fusion as a sustainable energy source has led to the development of multiple approaches, each with its own challenges and methodologies. Among these, two primary techniques dominate research and development: Magnetic Confinement Fusion (MCF) and Inertial Confinement Fusion (ICF).

MCF relies on strong magnetic fields to confine hot plasma within toroidal devices such as tokamaks and stellarators. Tokamaks are the most extensively studied, employing superconducting magnets to stabilize plasma and minimize disruptions [Ding, 2024]. Stellarators, by contrast, use complex, twisted external magnetic coils to create stable fields without relying on an induced plasma current, thus avoiding the instabilities associated with tokamaks and eliminating the need for continuous current drive [Sharma, 2024]. Meanwhile, Inertial Confinement Fusion (ICF) utilizes powerful lasers or ion beams to compress and heat small fuel pellets to conditions necessary for fusion. As the laser-driven approach aligns closely with the topic of this book, it is described in detail in the next section.

# 2.1 Laser-Driven Inertial Confinement Fusion

Laser Driven Inertial Confinement Fusion (ICF) is a method of achieving nuclear fusion by using lasers to compress and heat a tiny target of fusion fuel, typically deuterium and tritium, to the point where nuclear reactions occur. The principle behind ICF involves the rapid compression of the fuel target to extremely high densities and temperatures, creating conditions similar to those in the core of a star. The idea of using high-power lasers to compress the fuel target was first proposed in 1972 by Nuckolls et al. [Nuckolls, 1972]. This is achieved by directing multiple intense laser beams onto the surface of the target, causing the outer layer to ablate and generate a high-pressure shock wave that compresses the inner fuel to the point of fusion. The process relies on the inertia of the compressed fuel to confine the pressure long enough for the fusion reactions to occur, hence the term "inertial confinement." The goal is to reach a state where the energy generated by fusion reactions surpasses the energy required to initiate them - a condition known as ignition. This milestone was first achieved in 2022 when the National Ignition Facility (NIF) successfully attained scientific energy breakeven [ICF, 2024]. Figure 33.1 shows a three-orders-of-magnitude increase in target gain achieved over 11 years through experiments and modelling, from initial 50/50 DT implosions in 2011 to surpassing Lawson's criterion in 2021.

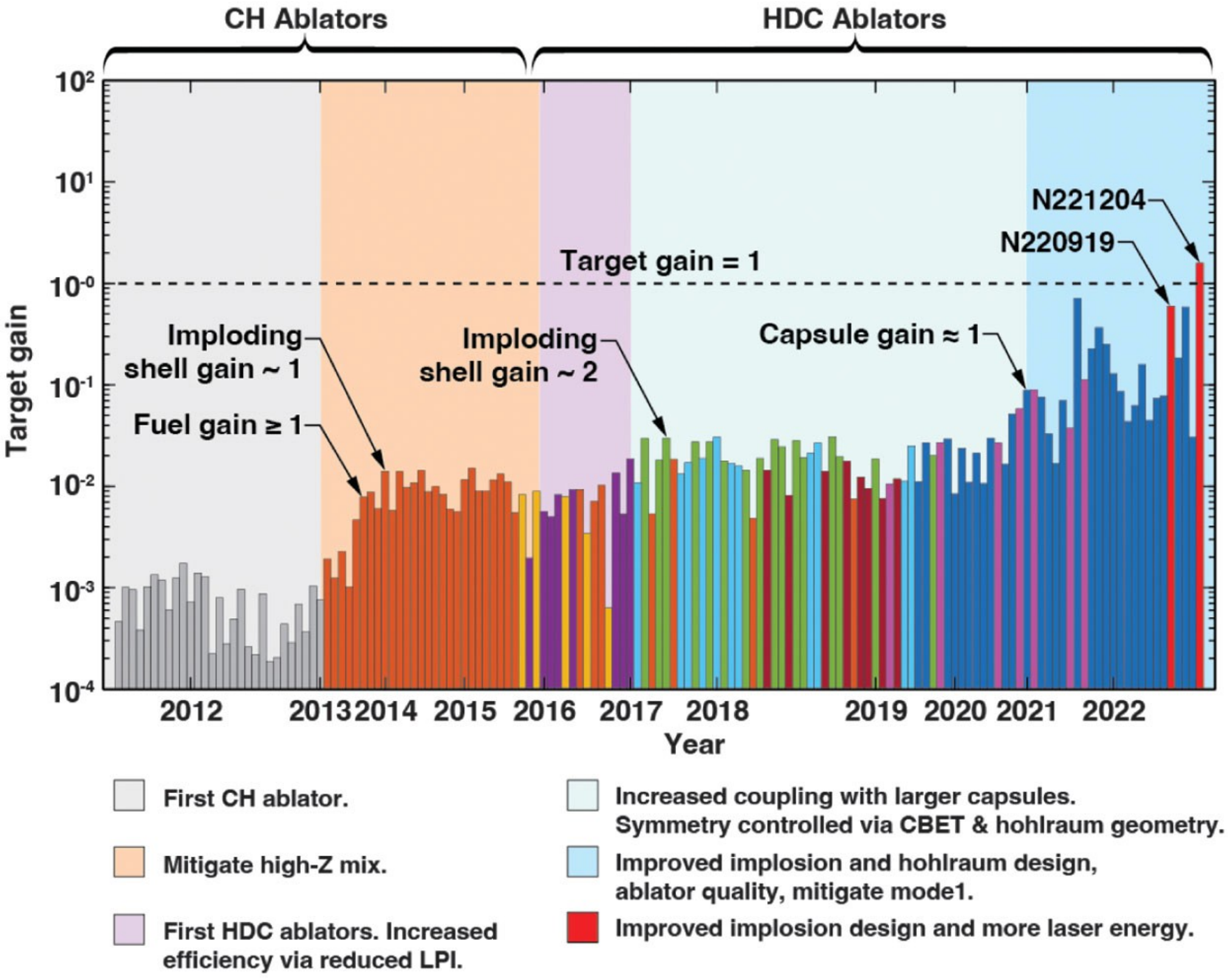

Figure 33.1: Target gain plotted against calendar date, where horizontal axis labels indicate the start of each calendar year; the colours of the narrow bars differentiate implosion designs, and the dashed horizontal line marks a target gain of 1, corresponding to the ignition criteria defined by the National Academy of Sciences (NAS). Figure reprinted from [ICF, 2024], H. Abu-Shawareb, R. Acree, P. Adams, J. Adams, B. Addis, R. Aden, P. Adrian, B.B. Afeyan, A.B. Zylstra, et al., Achievement of Target Gain Larger than Unity in an Inertial Fusion Experiment. Phys. Rev. Lett. 132, 065102 (2024), Copyright 2024, under Creative Commons Attribution 4.0 International License.

Deuterium-tritium (DT) is mainly used as a fuel target because it has the highest fusion cross-section (probability that two nuclei will successfully undergo fusion when they collide) at achievable temperatures, making ignition more efficient with current laser technologies. The DT reaction releases 17.59 MeV per fusion event, with energetic neutrons that facilitate energy extraction [Tahir, 1997]. Additionally, DT fuel benefits from efficient alpha-particle heating, which helps sustain the fusion process [Hurricane, 2016]. Despite challenges related to tritium's radioactivity and limited natural availability, its superior ignition efficiency makes it the preferred choice for inertial confinement fusion (ICF).

To compress DT fuel to the extreme densities and temperatures required for fusion ignition, ICF employs two principal techniques:

**Indirect-Drive ICF:** In this method, laser energy is not directly applied to the fuel pellet but is instead converted into x-rays within a hohlraum, a metal enclosure surrounding the fuel capsule (see Figure 33.2, left). These x-rays uniformly irradiate the target, driving an efficient implosion [Lindl, 1995]. Recent research has explored enhancements such as magnetized hohlraums to improve energy coupling and implosion symmetry [Strozzi, 2024].

**Direct-Drive ICF:** Unlike indirect-drive, this approach eliminates the hohlraum and directly targets the fuel pellet with multiple high-power laser beams (see Figure 33.2, right). While it offers potentially greater efficiency, challenges such as laser imprint instabilities (caused by fluctuations in beam intensity) must be mitigated. Advanced hybrid configurations, including two-sided ultraviolet laser irradiation, are being explored to improve implosion symmetry [Thomas, 2024].

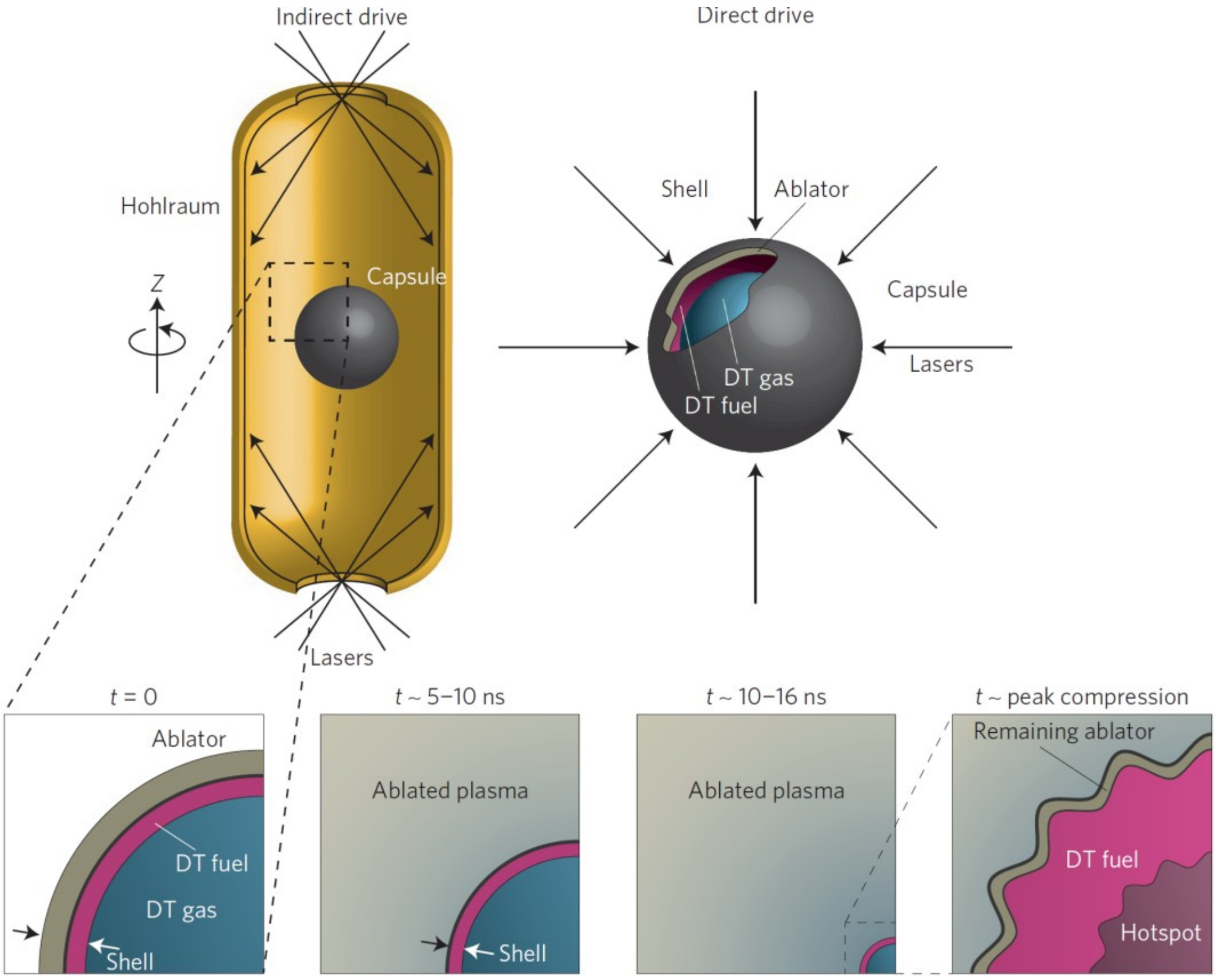

Figure 33.2: The conventional targets employed in laser-driven inertial confinement fusion (ICF) are categorized as either indirectly driven (upper left) or directly driven (upper right). In both circumstances, a spherical capsule is established at time tD0, featuring a layer of deuterium-tritium (DT) fuel on its internal surface. As the external surface of the capsule absorbs energy and undergoes ablation, the resultant pressure propels the remaining shell of ablator and DT fuel inward, resulting in an implosive event. Upon reaching approximately one-fifth of its initial radius, the shell attains a velocity of several hundreds of kilometers per second. By the moment the implosion achieves its minimum radius, a hotspot composed of DT has emerged, encircled by colder and denser DT fuel [Betti, 2016]. Figure reprinted from [Betti, 2016], R. Betti, O.A. Hurricane, Inertial-confinement fusion with lasers, Nature Phys 12, 435–448 (2016), Copyright 2016, with permission from Springer Nature.

## 2.2 Key Experimental Facilities for High-Energy ICF Research

Several experimental facilities worldwide utilize the mentioned ICF techniques to advance the understanding of high-energy-density physics. These installations represent critical infrastructures that enable researchers to experimentally explore the extreme conditions required for achieving nuclear fusion on Earth. Each facility typically employs specialized configurations and technologies tailored to distinct research goals. For instance, the Laser Mega Joule (LMJ) facility in France uses a sophisticated indirect drive method, harnessing 88 large-aperture laser beams to produce around 1.3 megajoules of ultraviolet laser energy that irradiates a target within a vacuum chamber [Neauport, 2024]. Similarly, Japan's Fast Ignition Realization Experiment (FIREX) employs an innovative two-step direct-drive fast ignition strategy, combining the compression capabilities of the GEKKO XII laser with the intense ignition pulse from the LFEX petawatt laser to improve energy efficiency [Fujioka, 2016]. In the United States, the Omega Laser Facility at the University of Rochester offers flexible experimental setups for both direct- and indirect-drive methodologies, featuring 60 beams capable of delivering around 30 kJ of ultraviolet laser energy [Waxer, 2005]. China's Shenguang-III (SG-III) facility, another key player, employs 48 laser beams capable of delivering approximately 200 kJ of ultraviolet energy, mainly supporting indirect-drive fusion experiments along with plasma physics and nuclear research, thus forming the technological and scientific groundwork for future, larger-scale fusion projects [Jiang, 2018]. The most powerful and extensively used facility globally is the National Ignition Facility (NIF) in the USA, capable of delivering an unprecedented 1.8 MJ and 500 TW of power using its sophisticated 192-beam, frequency-tripled neodymium-glass laser system operating at 351 nm [Di Nicola, 2018]. By carefully optimizing the laser parameters, NIF (on December 5, 2022), achieved a historic breakthrough in fusion ignition, where a controlled fusion reaction generated 3.15 MJ of energy, surpassing the 2.05 MJ of laser energy input and exceeding the scientific breakeven point for the first time [ICF, 2024].

## 2.3 Crucial Laser Parameters

To achieve ignition and high fusion energy yield requires precise control of laser parameters, including wavelength, pulse duration, intensity, energy, and beam uniformity. The following paragraphs provide an overview of each parameter, discussing their roles and impacts on fusion performance in ICF experiments.

**Laser intensity**: the power delivered per unit area ($W/cm^2$) dictates the strength of the ablation pressure and the rate of plasma blowoff from the target surface. Higher intensities produce higher ablation pressures via the rocket effect, which are needed to compress the fuel to fusion conditions. For example, typical laser intensities in the range of $10^{14}$ - $10^{15}$ $W/cm^2$ are used to drive the implosion of a capsule containing fuel in ICF, generating pressures of around 100 megabars required to achieve ignition conditions, including sufficient areal density, hotspot temperature, and confinement time, as stipulated by the Lawson criterion [Campbell, 1999; Lawson, 1957]. However, intensity cannot be increased arbitrarily without consequences. Above certain intensity thresholds, laser-plasma instabilities (LPI) such as stimulated Raman scattering (SRS), stimulated Brillouin scattering (SBS), and two-plasmon decay can grow in the plasma corona. These instabilities divert laser energy away from the target (via scattered light) and can accelerate electrons to high energies that penetrate and preheat the fuel (reducing compression efficiency) [Wasser, 2024]. There, careful management of the laser intensity is crucial to maximize the pressure on the target while minimizing energy loss to unwanted plasma waves.

The total **laser energy** sets fundamental limits on achievable compression and fusion yields. Modern facilities such as the National Ignition Facility (NIF) deliver megajoule-scale energies to meet the energy threshold required for ignition, typically around 1–2 MJ. Higher input energies facilitate higher implosion velocities and densities, necessary for ignition, but must be carefully controlled to avoid instability growth and energy loss due to nonlinear effects or optical damage [Hurricane, 2024].

**Wavelength** of the laser produced radiation also plays a significant role. The ICF community has shown a preference for utilizing shorter wavelengths for compression pulses due to enhanced laser-target coupling, increased ablation pressure for a specific intensity, and minimized laser plasma instabilities. Notably, the dominant laser light absorption mechanism of the generated plasma is collisional absorption, which is more efficient at shorter wavelengths, as the laser can propagate to denser regions where the collision frequency is elevated. This enhanced absorption at higher densities ensures effective ablation pressure and target compression, which are crucial for efficient implosion performance in inertial confinement fusion. Consequently, current ICF research facilities, such as the NIF, OMEGA, and the Laser MegaJoule (LMJ), function at a wavelength of 351 nm, which corresponds to the third harmonic (3x) of the 1053 nm fundamental (1x) of Nd:glass lasers [Labaune, 2013; Wasser, 2024].

**Beam uniformity** directly affects the symmetry and stability of target implosions. Non-uniform laser beams produce hot spots, leading to uneven ablation rates and hydrodynamic instabilities, compromising target integrity and fusion yield. Advanced beam smoothing techniques, such as smoothing by spectral dispersion (SSD), distributed phase plates (DPPs), and polarization smoothing (significantly improve uniformity), reducing instability growth and enabling higher-density, symmetric implosions crucial for successful ICF experiments [Rothenberg, 1997].

# 3. Laser Technology in Solar Energy Harvesting

Numerous laser-based processes have been developed for improving the production and the energy conversion efficiency of solar cells and photovoltaic modules [Compaan, 2000; Abott, 2007; Bartolome, 2010; Dunsky, 2011; Carlson, 2012; Gupta, 2015; Wang, 2023; Jamaatisomarin, 2023]. This includes all generations of solar cells, as visualized and reprinted in Figure 33.3 from [Amrillah, 2023] for the most common large area solar cell generations. The figure provides typical cross-sectional stratigraphies of the involved materials, lists corresponding values of cell and module efficiencies obtained until 2023, and compares other key properties of the different technological approaches. Not considered in the figure is the second generation amorphous silicon (a-Si) thin film solar cell technology that was developed and laser-processed already during the 1980's by a Japanese company [Nakano, 1986a; Nakano, 1986b].

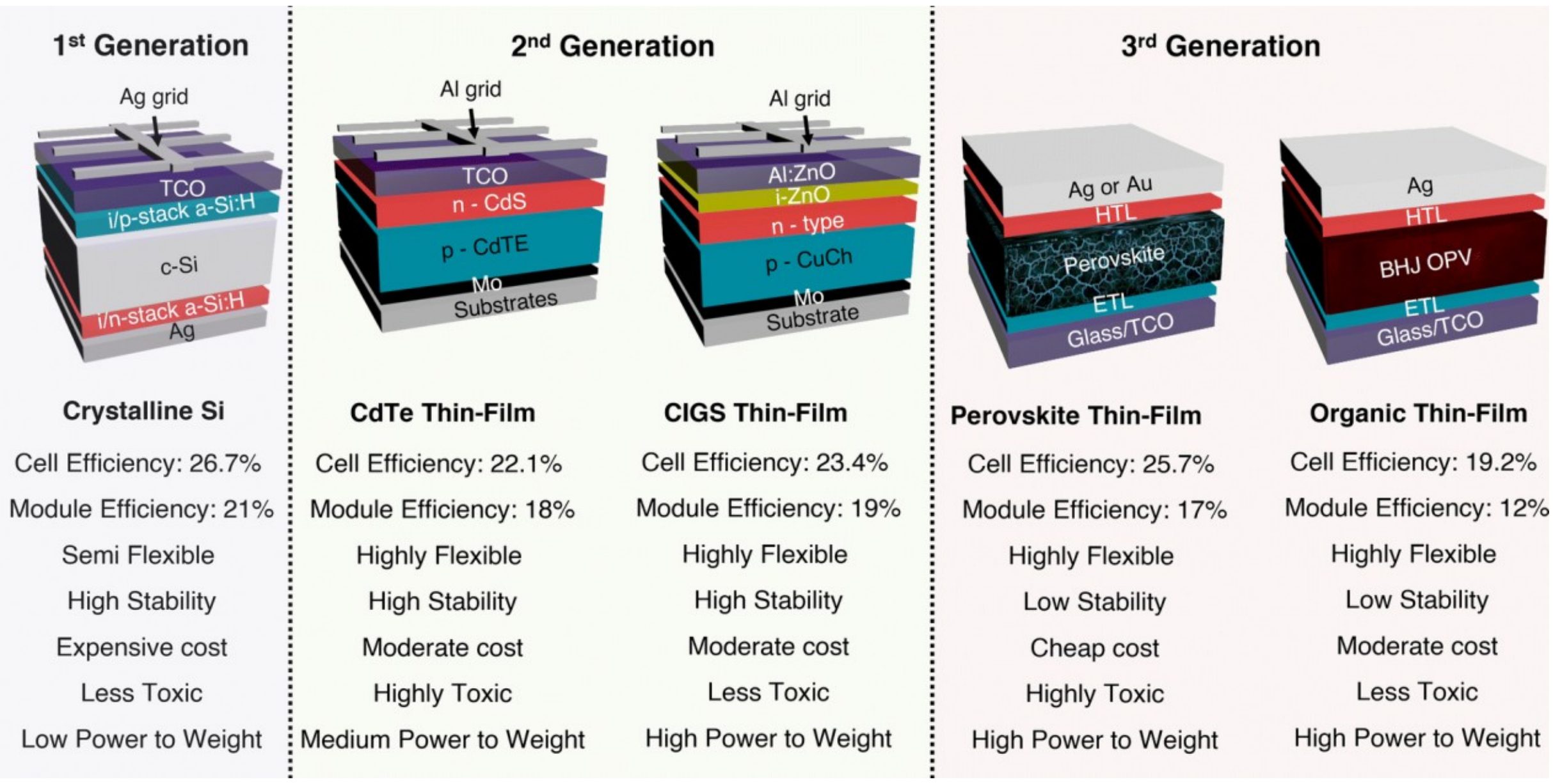

Figure 33.3: Photovoltaic wafer and thin film technologies [Amrillah, 2023]. The panels provide typical material stratigraphies along with a comparison of the three generations of solar cells. Used with permission of Royal Society of Chemistry, from [Amrillah, 2023], T. Amrillah, A. Prasetio, A.R. Supandi, D.H. Sidiq, F.S. Putra, M.A. Nugroho, Z. Salsabilla, R. Azmi, Environment-friendly copper-based chalcogenides thin film solar cells: Status and perspectives. Mater. Horiz. 10, 313 – 339 (2023), Copyright 2023; permission conveyed through Copyright Clearance Center, Inc.

## 3.1 Photovoltaic Cell Optimization with Lasers

For the first generation of bulk crystalline silicon (c-Si) wafer-based solar cells (for technology reviews see [Green, 2000; Saxena, 2024]), lasers were involved in numerous ways, e.g. for laser edge isolation, laser-grooved buried contacts, laser cutting of silicon ribbons, laser marking for cell or module identification, laser drilling for emitter-wrap-trough (EWT) contacts, laser-doped selective emitters, laser ablation of dielectric layers for localized contacts, laser defect/shunt repair, etc. [Abbott, 2007; Dunsky, 2011; Carlson, 2012; Bonse, 2013; Gupta, 2015]. Most of these laser-based processes and solar cell technologies were developed in research laboratories and companies in the USA, Germany, and Japan. Currently (2023/2024), more than 95% of the annually produced PV modules are based on monocrystalline silicon solar cells. Most of them (> 95%) are produced in Asia [Irena, 2024; Philipps, 2024a].

### 3.1.1 Laser Scribing

For saving material and reducing costs, the second and all following generations of large area solar cells were developed using thin film (TF) technologies for deposition of the solar absorber layer with a thickness tailored to the optical penetration depth of the solar radiation, as well as transparent conductive oxide (TCO) layers, or metallic back-contact layers (often made of molybdenum), while glass sheets are being used as cheap carrier material of the stacked, laterally structured and interconnected solar cell thin film stacks. The solar absorber layers typically consist of amorphous or (poly-)crystalline elemental group IV or III-V compound semiconductors, chalcogenides, perovskites, or organic materials. For manufacturing TF solar cells in a bottom-up strategy, the tailored electrical isolation of certain film segments, alternating with thin film deposition steps is required. These "scribing" steps can be conducted either mechanically by needles (with a limited precision and subjected to wear) or in a contact-less manner by precise laser processing [Compaan, 1995; NREL-Report, 1998; NREL-Report, 2000; Compaan, 2000; Bovatsek, 2010; Gecys, 2017; Wang, 2023; Bonse, 2023].

Figure 33.4a exemplifies a scheme as typical for a-Si TF solar cells, showing the geometry of three serially interconnected thin film solar cells [Bovatsek, 2010]. Three layers (L1: transparent contact film, L2: solar absorber (PV) film, L3: electric back contact film) are deposited on a glass carrier (substrate). The arrows indicate the direction of the current flow. Three vertical arrows mark the positions of electrically isolating "scribes" (P1, P2, P3), that are required between the thin film depositions for building up a series of interconnected cells: The P1-scribes singulate the TCO layer L1 into electrically isolated stripes on the glass substrate. Subsequently, the solar absorber layer L2 is deposited. In the following P2-scribing step, the L2-layer is completely removed – leaving behind an intact L1-layer on the glass. Next, the back electrode layer L3 is deposited. The final P3-scribes selectively remove both, the L2 and L3 layers (together), and finish the design of serially interconnected solar cells that can be assembled into a PV module.

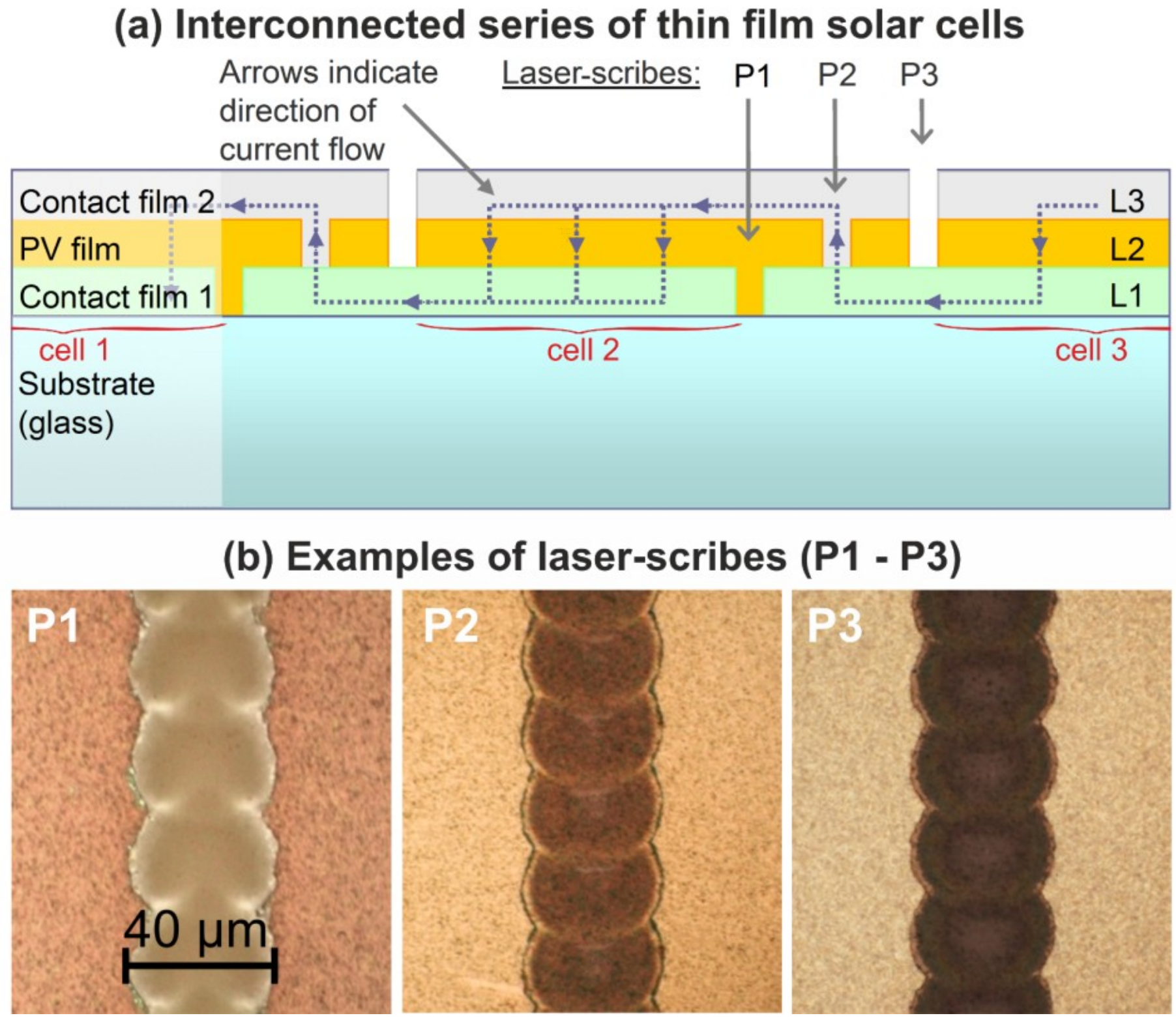

Figure 33.4: (a) Scheme showing the basic geometry and functioning of three serially interconnected thin film solar cells [Bovatsek, 2010]. (b) Top-view bright-field optical micrographs of a P1-scribe in a 650 nm thick $SnO_2$ layer on glass, a P2-scribe in a 220 nm thick a-Si layer on a $SnO_2$ film (675 nm) coated glass, and a P3-scribe in a 570 nm thick Al:ZnO/a-Si multi-layer on a $SnO_2$ film (660 nm) coated glass [Bovatsek, 2010]. All scribes were performed through the 3 mm thick glass substrate using a green ns-pulsed laser ($\lambda$ = 532 nm, $\tau_p$ = 13 ns, $f_{rep}$ = 20 to 60 kHz) along with a Galvanometer scanner, operated at 0.5 to 1.5 m/s scanning velocities. Reprinted from [Bovatsek, 2010], J. Bovatsek, A. Tamhankar, R.S. Patel, N.M. Bulgakova, J. Bonse, Thin film removal mechanisms in ns-laser processing of photovoltaic materials. Thin Solid Films 518, 2897 – 2904 (2010), Copyright 2009, with permission from Elsevier.

Figure 33.4b compiles three optical bright-field micrographs of P1-, P2-, and P3-lines laser-scribed with a focused ns-laser ($\lambda$ = 532 nm, $\tau_p$ = 13 ns, $f_{rep}$ = 20 to 60 kHz) along with a Galvanometer-scanner using scanning velocities between 0.5 and 1.5 m/s [Bovatsek, 2010]. Clean laser-scribes were obtained in all three cases. The spot overlap around 80% enabled an optimum processing time that ensured a scribe width between 30 µm and 40 µm.

For identifying the film removal mechanism acting during the laser-scribing, numerical simulations were performed for solving the *heat equation* under the boundary conditions (film stratigraphy and materials, laser excitation, processing parameters) given for the P2-scribe provided in Figure 33.4b above [Bovatsek, 2010]. The results were visualized in color-encoded spatio-temporal temperature maps displaying the depth-resolved temperatures vs. time, as provided in Figure 33.5a. It turned out that at the given laser fluences maximum temperatures of ~1350 K are obtained close to the a-Si/$SnO_2$ (TCO) interface. Since that value is slightly below the melting temperature of amorphous silicon (1420 K), and further supported by a thermo-elastic model, it was concluded that a laser-induced thermal stress (strain) related mechanism led to a (clean) delamination of the a-Si film from the underlying glass substrate, here [Bovatsek, 2010].

Figure 33.5b provides a generalized scheme that summarizes the possible processes that can be used for through-the-glass laser-scribing in thin film photovoltaics. The laser-induced film removal occurs in three steps: during step 1, the focused laser radiation is absorbed in the solar absorber layer L2 close to the glass-L2 interface. In step 2 thermal strain, melting, or even evaporation emerges at the interface, before in step 3 the disintegration of the film occurs via fracture, evaporation, etc. – eventually leading to its ejection and removal from the glass substrate [Bovatsek, 2010].

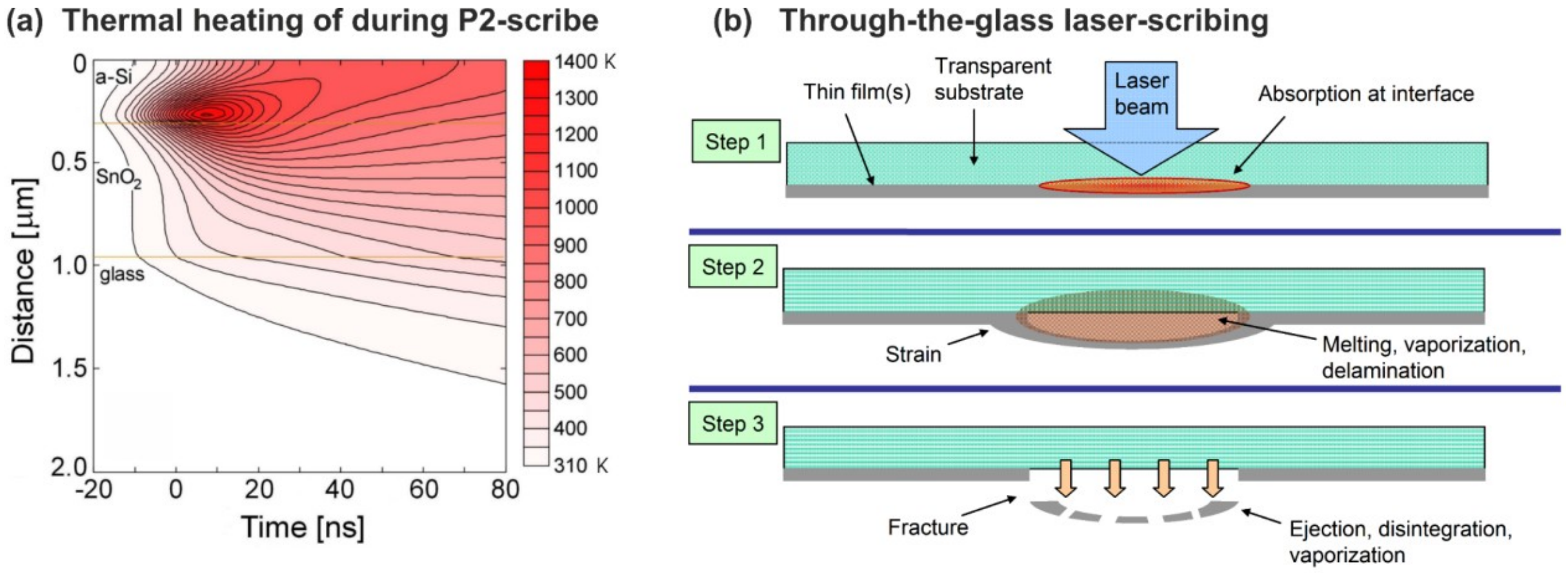

Figure 33.5: (a) Spatio-temporal temperature map obtained by numerically solving the heat equation for the boundary conditions given by the P2-scribe shown in Figure 33.4b. (b) Scheme illustrating thin film removal upon laser incidence through a transparent substrate. Reprinted from [Bovatsek, 2010], J. Bovatsek, A. Tamhankar, R.S. Patel, N.M. Bulgakova, J. Bonse, Thin film removal mechanisms in ns-laser processing of photovoltaic materials. Thin Solid Films 518, 2897 – 2904 (2010), Copyright 2009, with permission from Elsevier.

### 3.1.2 Laser Texturing

One important goal of photovoltaic research is to increase solar cell efficiency, aiming to reduce the cost of installed solar power, while simultaneously decreasing the area required per watt of energy produced. Efficiency improvements not only lower the overall costs of PV installations but also maximize the energy yield from limited space, a critical factor in densely populated or resource-constrained environments. Among the many and varied strategies for efficiency improvement, texturing the surfaces or interfaces in solar cells has emerged as an effective approach. Usually, the primary function of texturing is to increase the optical absorption by minimizing the loss of photons that would otherwise be reflected off the smooth surface of a standard solar cell. By creating a roughened or textured surface, light incident on the cell's top surface is more likely to be transmitted into the solar absorber material and at the same time refracted at oblique angles or scattered, which enhances the likelihood of photon absorption by increasing the optical path length. These phenomena are particularly beneficial for thin-film solar cells, where absorbing materials are limited in thickness, as well as for weak-absorbing materials, such as crystalline silicon in the NIR spectrum. It has also been reported that thin-film solar cells with textured interfaces can benefit from an enhanced charge carrier collection [Pascoe, 2016]

Various surface texturing techniques have been developed, including chemical etching, laser texturing, and mechanical abrasion, each offering unique advantages depending on the specific material and cell architecture. Traditionally, in monocrystalline silicon (mono-Si) solar cells anisotropic wet chemical etching in alkaline solution is used to selectively remove material at different rates depending on the crystallographic orientation, inducing the formation of pyramid-like texture with typical lateral resolution in the order of 5 – 20 µm. In case of multicrystalline Si (mc-Si), enhanced light trapping was achieved by etching in an acidic solution [Zielke, 2013]. However, wet chemical etching presents certain challenges, such as controlling etch uniformity across large wafers and managing waste from chemical processing. In contrast, laser texturing can provide significant benefits, as it is a one-step green-technology approach with the potential for high-throughput processing at marginal operation costs. Moreover, the lateral resolution could be improved toward the sub-micro range, enabling light trapping within the domain of physical optics.

In 1998, the group of Mazur at the Harvard University (USA) explored the manufacturing of so-called "black silicon" [Her, 1998] upon irradiating the silicon surface using Ti:sapphire femtosecond laser pulses in an $SF_6$ atmosphere. This kind of surface morphology is characterized by the presence of characteristic micro-cones formed in a self-ordered way upon fs-laser irradiation of silicon crystals at high laser fluences (a few J/cm$^2$) and with a large number of laser pulses per spot (typically hundreds) [Bonse, 2002]. This characteristic topography turns the visual appearance of silicon to an almost perfectly black surface. The approach of Mazur and co-workers using a sulfur-containing atmosphere during fs-laser treatment has the additional advantage of inducing the diffusion and incorporation of sulfur atoms into the silicon wafer, enabling localized doping, a process coined as "hyperdoping". In this direction, different research groups have demonstrated large decreases in reflectivity of silicon wafers by patterning different types of self-ordered laser-generated surface structures using ultrashort laser pulses in different controlled gas atmospheres and in air. Figure 33.6a shows the reflectance of such laser-treated wafers as published by various authors, compared to the polished surface and the conventional pyramidal texture obtained by etching mono-Si (without antireflective coatings). Binetti et al. have shown that laser texturing on mc-Si using a ps-laser source is an effective approach to reduce the reflectance down to 8% in the solar absorbing range of Si (approximately from 300 nm to 1100 nm), outperforming conventional techniques based on wet chemical etching [Binetti, 2016]. As observed in Figure 33.6b-d, the morphology of the laser-induced surface textures depended on the used wavelength and it tended to produce a higher concentration of microholes for the wavelengths of 532 nm and 1064 nm. Although the presence of these microholes might contribute to the reduction of the reflectance, the authors report that they lead to degradation of the device performance as they penetrate into the depletion region of the p-n junction. In case of irradiating the surface with UV laser light (355 nm), the silicon surface reflectance could be reduced down to 9% on average and as the structures were shallower, they did not reach the depletion region and thus did not compromise the electronic quality. Also, fs-pulsed laser radiation (800 nm) was employed for texturing the surface of mono-Si wafers and to reduce the reflectance (Figure 33.6a). Sharma et al. showed that by increasing the cumulated fluence (by varying pulse-to-pulse overlap and fluence per spot) from 8 J/cm$^2$ to 11 J/cm$^2$, the surface morphology evolved from spheroids with feature sizes of 1 μm to 2 μm (Figure 33.6e) to conical spikes with lateral size on the order of 10 μm (Figure 33.6f), respectively [Sharma, 2022]. Although the resulting reflectances induced by these microstructures were similar, the spheroids showed an overall smaller value probably caused by an increased light scattering triggered by the smaller feature sizes and denser concentration of microstructures. Using a similar laser source, Kontermann and co-workers [Kontermann, 2012] textured mono-Si at a high cumulated fluence of 1000 J/cm$^2$, achieving high-aspect ratio spikes with a conically-tapered shape (Figure 33.6g). This specific morphology allowed for an extreme light absorbance even extending towards the near-IR spectrum.

The irradiation of crystalline semiconductors with laser light might induce crystallographic defects and structural changes within the heat-affected zone, potentially reducing the electronic quality of Si and increasing the surface recombination of photogenerated charge carriers. To mitigate this inherent drawback of the laser microtexturing process, different strategies inherited from traditional silicon technology have been implemented to passivate surface defects [Drygała, 2016; Nayak, 2011; Radfar, 2018]. For instance, chemical etching may remove the defective surface layer. Alternatively, a post-process deposition of a thin oxide (e.g. $Al_2O_3$, $SiO_2$) or nitride (e.g. $SiN_x$) layer with a thickness of a few tens of nm neutralizes defect states and reduces surface recombination sites by saturating dangling bonds at the silicon surface. Typically, techniques such as atomic layer deposition (ALD) or plasma-enhanced chemical vapor deposition (PE-CVD) are used for this purpose. In addition, these thin layers serve as anti-reflective coatings, further depleting the solar cell reflectance.

As an example, Zielke et al. reported performance improvements of lab-scale mono-Si and mc-Si solar cells upon laser texturing using ns-pulses followed by etch cleaning of the laser process residues and deposition of oxides and nitrides passivation layers [Zielke, 2013]. The efficiency improvements were 8.7% and 7.2% (relative) in the mono-Si and mc-Si solar cells, compared to the flat reference cells, respectively. These boosts in performance were mainly driven by a better light absorption, allowing the cells to deliver higher photocurrents. Furthermore, the authors observed that the efficiencies of the laser textured cells were in line with the values of the solar cells featuring random pyramids produced by wet etching.

On the downside, laser texturing of c-Si for industrial-scale photovoltaics still lags behind conventional methods based on wet chemical etching, mainly due to the initial high capital costs for the laser equipment and relatively low processing times compared to batch etching of large (e.g., 6 inches) wafers. In addition, after the laser texturing step, normally an etching procedure is still required to remove the ablation slag and smooth the surface [Sreejith, 2022].

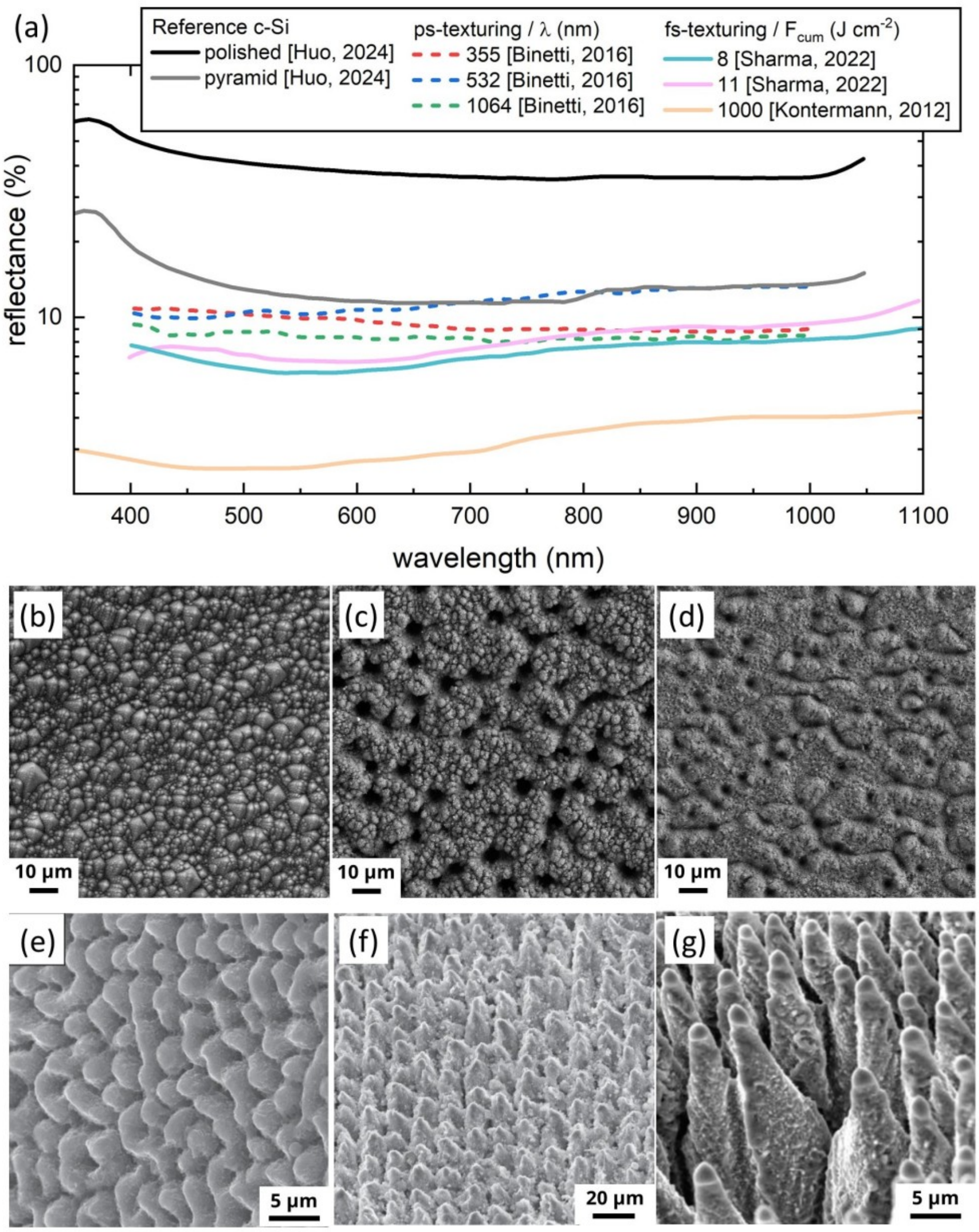

Figure 33.6: (a) Reflectance of laser-irradiated silicon surfaces using ultrashort pulses. SEM images of textures produced on c-Si by ps-pulses at a wavelength of (b) 355 nm, (c) 532 nm, and (d) 1064 nm and by fs-pulses at a cumulated fluence of (e) 8 J/cm$^2$, (f) 11 J/cm$^2$, and (g) 1000 J/cm$^2$. See text for details. Note the different magnifications in the micrographs. Figures (b) to (b) adapted from [Binetti, 2016] with permission from Elsevier. Figure (e) and (f) reproduced from [Sharma, 2022], Sharma, A.; Marla, D.; Joshi, S. S.; Bathe, R.: A study of femtosecond laser processed microtextures on silicon wafers to enhance optical absorption. Lasers Manuf. Mater. Process. 9, 277–291 (2022), with permission from Springer Nature. Copyright 2022. Figure (g) reprinted from [Kontermann, 2012], Kontermann, S.; Gimpel, T.; Baumann, A. L.; Guenther, K.-M.; Schade, W.: Laser processed black silicon for photovoltaic applications. Energy Procedia 27, 390–395 (2012), Copyright 2012, with permission from Elsevier.

In thin film solar cells, the absorber material is characterized by a thickness on the order of 1 µm (or considerably thinner) and a total thickness of the solar cells on the order of a few µm. Thus, material-saving thin film solar cells have the potential to be very lightweight and they can be deposited on flexible substrates, which can expand their deployment into wearables or building integrated photovoltaics even with aesthetic value. Furthermore, typically the fabrication methods used for thin film solar cells are considerably more cost-effective than those required for wafer-based photovoltaics. Normally, the solar absorber and functional layers, such as the metal contacts, are deposited sequentially by physical vapor deposition (PVD) methods or by solution-based processes, such as spin- or slot-die coating. However, due to the reduced absorber thickness, efficient sunlight absorption can be challenging. In this context, texturing the front surface or one of the internal interfaces has been explored by many research groups in order to promote a better light trapping. Particularly, many research studies have been focused on laser texturing the first functional layer deposited on the substrate, i.e., the front electrode, which is typically made of a transparent conductive oxide (TCO), such as the mixture of $In_2O_3$ and $SnO_2$ (ITO) or fluorine-doped $SnO_2$ (FTO). If the surface texture is periodic and its periodicity is comparable to the sunlight wavelength spectrum range (i.e., ~1 µm), the patterned TCO can act as a transmissive diffraction grating, splitting the incoming light into well-defined angles. This can thus lead to longer optical paths within the absorber material and enhanced absorption. If the surface texture exhibits irregular near- or sub-wavelength features, optical scattering effects can improve the light management similarly via extending the optical path lengths within the cell stratigraphy and via refractive index matching effects [Knüttel, 2013a], see Figure 33.7.

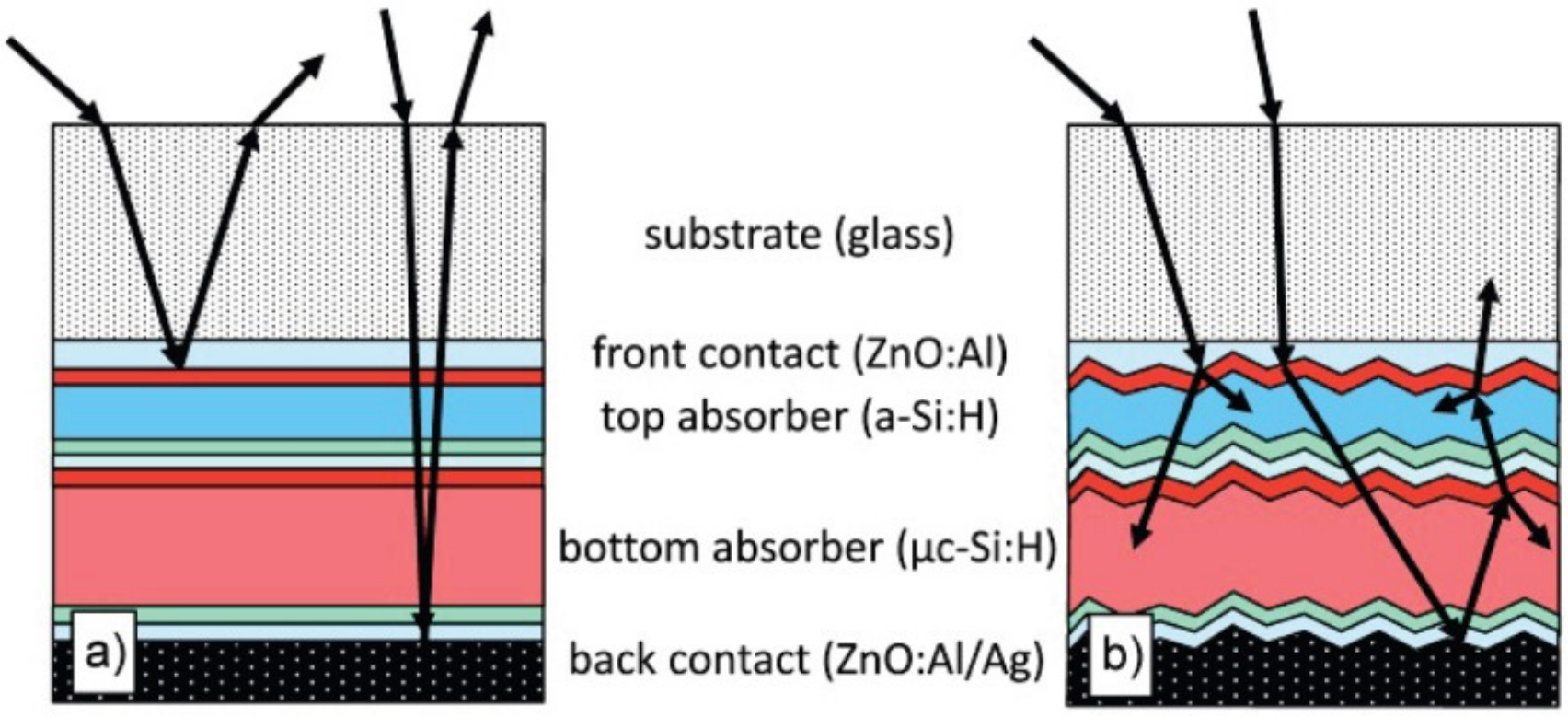

Figure 33.7: (a) Scheme illustrating the light management in thin film silicon solar cells (a) with flat interfaces and (b) with textured interfaces. The textured film interfaces improve the cell efficiencies by extending the optical path lengths within the cell stratigraphy and via refractive index matching. Reproduced from Ref. [Knüttel, 2013a] with permission from Japan Laser Processing Society.

Using the direct laser writing (DLW) technique, arbitrary texture shapes can be patterned with a lateral resolution corresponding to the spot size, which can be typically in the range from 10 to 30 µm. Considering that the TCOs used in thin film photovoltaics have typically thicknesses on the order of a few hundreds of nm, the material ablation rate must be very well controlled so that the remaining TCO material keeps a good enough electrical conductivity to avoid resistive losses. In this direction, Canteli et al. patterned ZnO:Al (AZO) with line-like and dot-like textures using a ns-pulsed laser working in the third harmonic (355 nm) [Canteli, 2019]. As the obtained texture periodicity was ~10 µm, the TCOs worked as diffraction gratings increasing the haze (ratio between diffuse and direct transmittances) to values up to ~30% in the visible spectrum. The authors fabricated solar cells based on amorphous silicon (a-Si) and reported a relative increase in the device photocurrent of 14% compared to the flat reference cell. Following the same strategy, the optical path inside the absorber material could be further increased by patterning diffraction gratings on the TCOs with shorter periodicities, which, according to the well-known grating equation, yield larger diffraction angles. A more suitable technology to achieve this is the direct laser interference patterning (DLIP), which offers the possibility to produce textures with a periodicity even smaller than the laser wavelength. Several studies have been reported employing this technique for patterning periodic 1D or 2D gratings on different TCOs. A comprehensive morphological analysis of DLIP-treated AZO films using ns-pulses was conducted in [Eckhardt, 2013].

Figure 33.8 illustrates representative examples of (a) line-like and (b) hexagonal pillar-like textures reported in their study. The authors identified thermally induced surface tension gradients as the primary mechanism responsible for the displacement of molten material from the interference maximum to minimum positions, consistent with observations made during ns-DLIP processing of ZnO:B [Berger, 2016]. During the cooling phase following the laser-induced melting, the material crystallized into large grains, accompanied by a noticeable distribution of cracks, aligning with the findings in [Parellada-Monreal, 2018; Knüttel, 2013b]. In contrast, the formation of textures on AZO films using ps-pulsed DLIP exhibited a significant contribution from ablation, with a reduced presence of molten material compared to ns-pulsed processing. This difference is attributed to the shorter thermal diffusion length, below 20 nm [Ring, 2015], in the ps-regime compared to the ~1 μm thermal diffusion length in the ns-regime. The shorter thermal diffusion length results in highly localized heating and temperatures exceeding the vaporization point of AZO.

Using ps-DLIP 1D gratings with a spatial period of 860 nm were produced on AZO films and then hydrogenated a-Si/microcrystalline silicon (μc-Si) tandem solar cells were produced on top. Due to a better photon collection driven by the improved haze of the TCO, a relative power conversion efficiency improvement of 8% compared to the unstructured device was achieved [Ring, 2015]. Also 2D gratings were fabricated by DLIP combining four coherent beams from a ps-pulsed laser source (355 nm). Namely, Heffner et al. reported the irradiation of FTO-coated substrates with different fluence levels to produce dot-like textures with a spatial period of 700 nm [Heffner, 2025]. The SEM images (Figure 33.8c) revealed that the ablation took place along the grain boundaries without melting the remaining crystallites of the film and without yielding significant redeposited debris. As a result, no significant morphological changes, such as re-crystallization around the interference maxima positions, were observed strongly suggesting that cold ablation occurred. This points to a rapid and localized thermal energy increase within individual grains leading to their localized ablation, while the temperatures of neighboring grains remained even below the melting point, limiting the circularity of the craters to the boundaries of the grains. The DLIP-treated TCOs were then used to fabricate organic-inorganic hybrid perovskite solar cells with an enhanced average efficiency of 19.4% compared to the average efficiency of 18% of the control devices. This increase was attributed to the improved optical properties of the TCO, i.e. reduced global reflectance and improved haze.

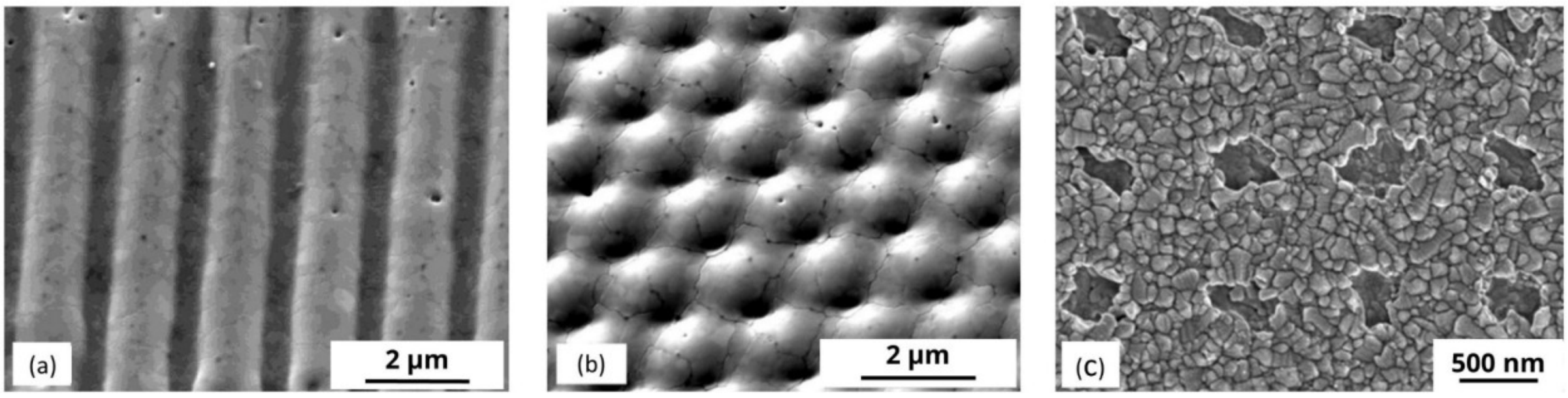

Figure 33.8: SEM images showing the morphology of DLIP textures on TCO electrode layers produced by DLIP, featuring (a) line-like, (b) hexagonal pillar-like, and (c) dot-like patterns. See the text for details. Figures (a) and (b) reprinted from [Eckhardt, 2013], Eckhardt, S.; Sachse, C.; Lasagni, A. F.: Light management in transparent conducting oxides by direct fabrication of periodic surface arrays. Phys. Procedia 41, 552–557 (2013), Copyright 2013, with permission from Elsevier. Figure (c) adapted from [Heffner, 2025], Heffner, H.; Du, Y.; Shilovskikh, V.; Taretto, K.; Wrzesińska-Lashkova, A.; Soldera, M.; Lasagni, A. F.; Vaynzof, Y.: Direct laser interference patterning of fluorine-doped tin oxide as a pathway to higher efficiency in perovskite solar cells. Adv. Funct. Mater. 35, 2415126 (2025), under Creative Commons Attribution 4.0 International License (https://doi.org/10.1002/adfm.202415126).

## 3.2 Concentrated Solar Power Using Lasers

Another promising concept for improving thin film solar cells is based on micro-concentrator photovoltaics (micro-CPV), see Figure 33.9. Generally, concentrator photovoltaics (CPV) uses an optical system to collect and harvest solar radiation and concentrates it on a smaller solar cell. From the material consumption point of view, this approach allows to save expensive or even scarce materials when compared to flat panel photovoltaics. The optical system collects the sunlight and confines it on the smaller solar cell with a higher power density, which, in turn, can increase the efficiency of the solar cell. However, CPV usually requires sun-tracking, is less sensitive to diffuse light collection, and is, thus, not suitable for rooftop applications.

Micro-CPV relies on a bottom-up approach, where sub-mm laterally sized thin film solar cells are site-controlled manufactured on a suitable heat-dissipating substrate, e.g., in a regular two-dimensional array of $N$ cells in both orthogonal directions. Each cell consists of a solar micro-absorber that is sandwiched between suitable electrode layers. In front of the cells, a tailored microlens array (sharing the same 2D geometry) is fixed that collects the sunlight and concentrates it via focusing onto the individual micro-absorbers (Figure 33.9a). Compared to ordinary CPV, the micro-CPV concept generally exhibits several advantages, such as reduced material consumption during manufacturing, cost reduction, increased optical efficiency, and improved compactness of the assembled modules [Alves, 2020], see Figure 33.9b. Due to its improved heat dissipation capabilities, the operation occurs at lower cell temperatures with beneficial effects on series resistance losses, thermal stress, and other degradation properties.

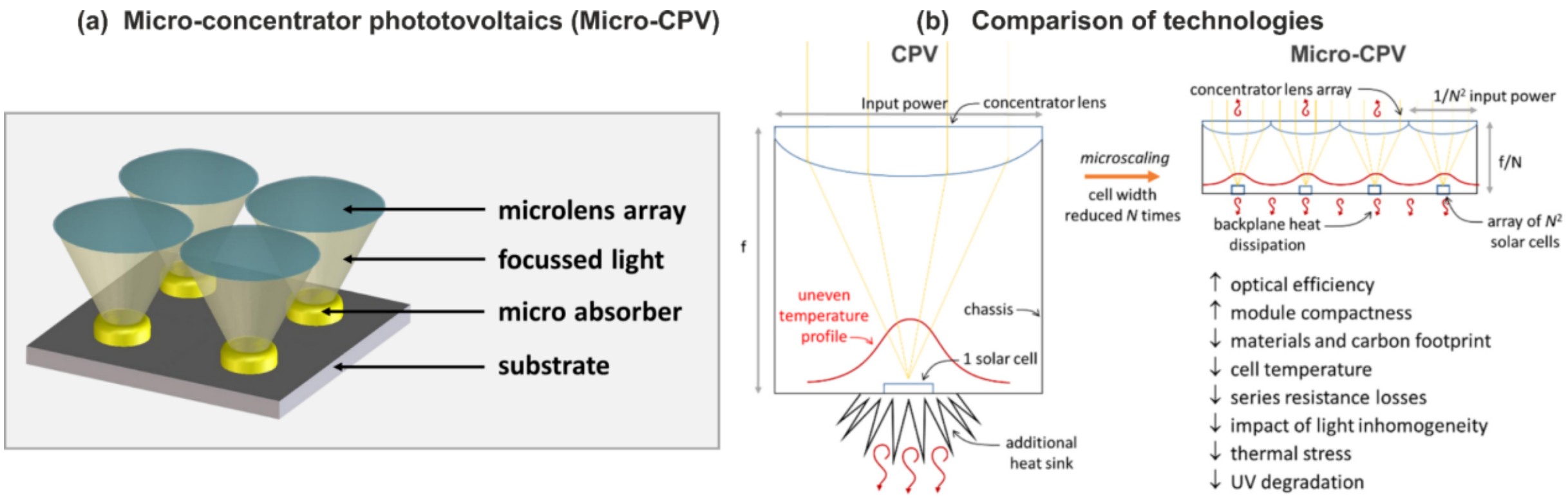

Figure 33.9: (a) Scheme of the micro-concentrator solar cell concept [Ringleb, 2018]. (b) Scheme comparing the concentrator photovoltaics (CPV, left column) with micro-concentrator photovoltaics (Micro-CPV, right column) [Alves, 2020]. Figure (a) reproduced from [Ringleb, 2018], F. Ringleb, S. Andree, B. Heidmann, J. Bonse, K. Eylers, O. Ernst, T. Boeck, M. Schmid, J. Krüger, Femtosecond laser-assisted fabrication of chalcopyrite micro-concentrator photovoltaics. Beilstein J. Nanotechn. 9, 3025 – 3038 (2018), under Creative Commons Attribution 4.0 International License (https://doi.org/10.3762/bjnano.9.281). Figure (b) reproduced from [Alves, 2020], M. Alves, A. Pérez-Rodríguez, P.J. Dale, C. Domínguez, S. Sadewasser, Thin-film micro-concentrator solar cells. J. Phys. Energy 2, 012001 (2020), under Creative Commons Attribution 3.0 International License (https://doi.org/10.1088/2515-7655/ab4289).

Several different concepts were proposed for the bottom-up manufacturing of thin film micro-concentrator solar cells, including area-selective physical vapor deposition (PVD), electrodeposition, and site-selective laser-based processing [Alves, 2020]. A brief review on laser-based processing strategies for micro-CPV based on the CIGSe ($Cu(In,Ga)Se_2$) or CISe ($CuInSe_2$) material systems can be found in [Ringleb, 2018], from which two different approaches are outlined in the following.

In the **laser-induced nucleation approach** (Figure 33.10a), a fs-laser was used for site-selective laser micro-structuring (surface roughening) of a molybdenum (Mo) layer on a glass substrate in the desired 2D-array pattern. Subsequent PVD allowed the selective nucleation of micrometer-sized indium (In) precursor islands at the laser irradiated spots (Figure 33.10b) [Ringleb, 2016; Ringleb, 2017]. Such In islands were subsequently coated by Cu and converted into CISe by a selenization process, after which excess Cu-Se was removed by a KCN chemical etching step. In this way, CISe or CIGSe micro-absorbers were successfully manufactured (Figure 33.10c). To isolate the solar cells, a resist layer was spin coated, leaving the upper part of the micro-absorber islands uncovered. Following coating by a CdS buffer layer and a transparent conductive oxide (ZnO) allowed the assembly into fully functional micro-solar cells [Heidmann, 2017].

The **laser-induced forward transfer (LIFT) approach** (Figure 33.10d), used the well-established LIFT technique (for a review see [Serra, 2019]) for site-selectively transferring precursor materials (In, Cu, Ga) from a donor layer onto the Mo-coated back contact on the glass substrate (acceptor) [Andree, 2017]. The optical micrograph presented in Figure 33.10e exemplifies such a LIFTed Cu-In deposit on a Mo acceptor layer. The LIFTed Cu-In precursors were then successfully processed into CIGSe micro-absorbers (Figure 33.10f) and assembled into micro-solar cells [Heidmann, 2018].

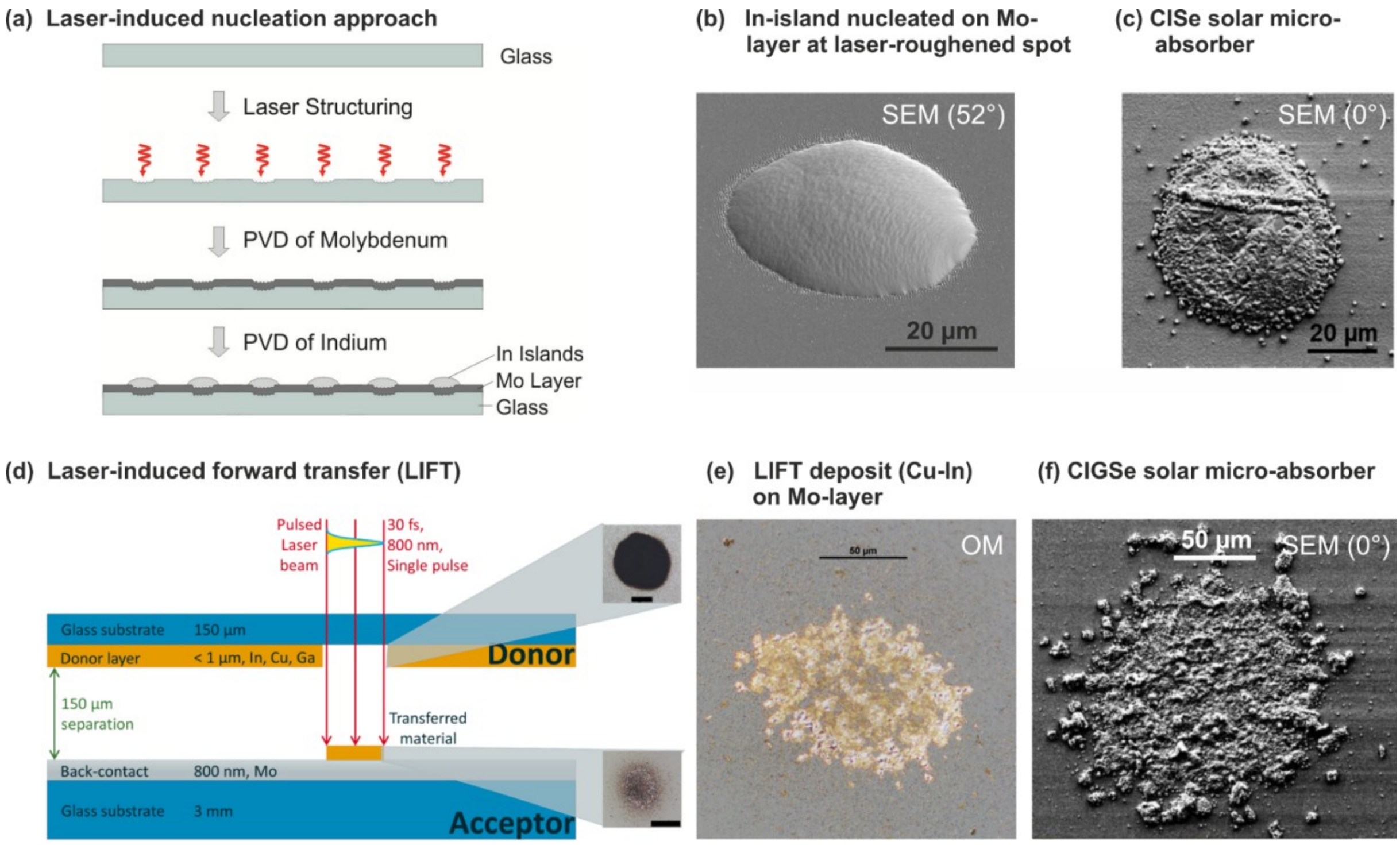

Figure 33.10: Two different laser-assisted bottom-up strategies for the fabrication of solar micro-absorbers. (a) Scheme of the laser-induced nucleation approach for site-selective surface roughening of the Mo-layer on a glass substrate [Ringleb, 2018]. (b) Tilted-view SEM micrograph of a single indium precursor island, nucleated and grown by PVD on the laser-structured Mo-layer [Ringleb, 2018]. (c) Top-view SEM micrograph of a single CISe micro-absorber post-processed from the In precursor island [Ringleb, 2018]. (d) Scheme of the laser-induced forward transfer (LIFT) approach for site-controlled donor material deposition on the acceptor [Ringleb, 2018] (e) Optical micrograph of a Cu-In-precursor LIFTed onto a Mo-layer on glass. (f) Top-view SEM micrograph of a single CISe micro-absorber post-processed from the LIFTed Cu-In-precursor island [Heidmann, 2018]. Figures (a-e) reproduced from [Ringleb, 2018], F. Ringleb, S. Andree, B. Heidmann, J. Bonse, K. Eylers, O. Ernst, T. Boeck, M. Schmid, J. Krüger, Femtosecond laser-assisted fabrication of chalcopyrite micro-concentrator photovoltaics. Beilstein J. Nanotechn. 9, 3025 – 3038 (2018), under Creative Commons Attribution 4.0 International License (https://doi.org/10.3762/bjnano.9.281). Figure (f) reprinted from [Heidmann, 2018], B. Heidmann, S. Andree, S. Levcenko, T. Unold, D. Abou-Ras, N. Schäfer, J. Bonse, J. Krüger, M. Schmid, Fabrication of regularly arranged chalcopyrite micro solar cells via femtosecond laser-induced forward transfer for concentrator application. ACS Appl. Energy Mater. 1, 27 – 31 (2018), Copyright 2017, with permission from Elsevier.

Figure 33.11 compiles some results obtained for CISe micro-absorber solar cells manufactured by the laser-induced nucleation approach. The left panel (Figure 33.11a) displays post-colorized tilted-view SEM micrograph of a cross-section of a CISe micro-absorber after processing into a solar cell. The micro-crystalline CISe absorber island is clearly visible and sandwiched between the covering CdTe buffer and TCO layers and the underlying Mo back contact layer [Ringleb, 2018]. The absorber island is laterally isolated by a photoresist layer. The middle panel (Figure 33.11b) provides typically measured *I-V* curves of an assembly of ~100 parallel connected CISe micro cells with (red curve) and without (black curve) illumination at one AM1.5 sun standard [Heidmann, 2017]. The expected behavior of solar cells is confirmed. The right panel (Figure 33.11c) shows two curves characterizing the performance of the resulting CISe micro-absorber solar cells [Ringleb, 2018], i.e., measured data for the open circuit voltage $V_{OC}$ and the cell efficiency, both as a function of the light concentration ratio $C$.

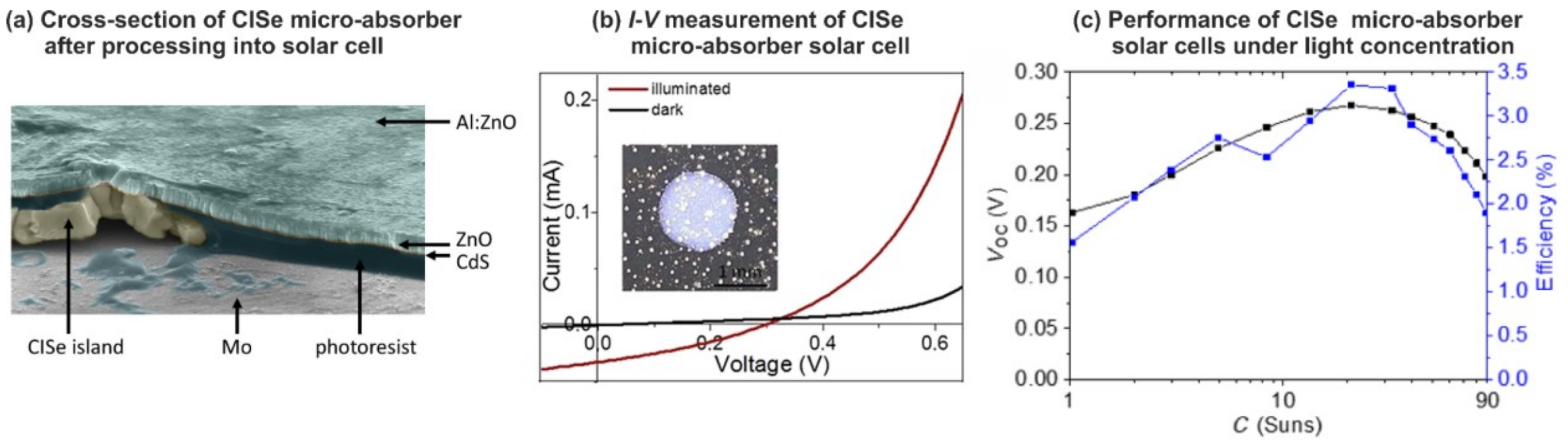

Figure 33.11: Compilation of results obtained for micro-absorber solar cells fabricated by the laser-assisted nucleation approach. (a) Tilted-view SEM image of a cross section of a CISe micro absorber island after processing to a micro cell [Ringleb, 2018]. The different materials were artificially post-colorized to enhance their visibility. The height of the polycrystalline CISe micro-absorber is ≈ 1 µm. (b) *I-V* measurement of micro solar cells (dark curve in black and illuminated curve in red) suing a solar simulator under AM1.5 standard one sun illumination. Inset: Optical micrograph of the measured sample area including about 100 micro-absorbers (white dots) [Heidmann, 2017]. (c) Performance of CISe micro-absorber solar cells (open circuit voltage VOC and cell efficiency) under concentrated light illumination [Ringleb, 2018]. Figure (a, c) reproduced from [Ringleb, 2018], F. Ringleb, S. Andree, B. Heidmann, J. Bonse, K. Eylers, O. Ernst, T. Boeck, M. Schmid, J. Krüger, Femtosecond laser-assisted fabrication of chalcopyrite micro-concentrator photovoltaics. Beilstein J. Nanotechn. 9, 3025 – 3038 (2018), under Creative Commons Attribution 4.0 International License (https://doi.org/10.3762/bjnano.9.281). Figure (b) reprinted from [Heidmann, 2017], B. Heidmann, F. Ringleb, K. Eylers, S. Levcenco, J. Bonse, S. Andree, J. Krüger, T. Unold, T. Boeck, M.C. Lux-Steiner, M. Schmid. Local growth of CuInSe2 micro solar cells for concentrator application. Mater. Today Energy 6, 238 – 247 (2017), Copyright 2017, with permission from Elsevier.

Although the electrical efficiency of these laser-manufactured micro solar cells is not yet comparable to that of large-area flat panel thin film CISe solar cell, the still very young technology exhibits a promising development along with the unique material-saving benefit [Alves, 2020].

## 3.3 Laser-Assisted Recycling of Solar Modules

Typical Si solar modules are made up of many components containing very different materials. The Si wafers must be electrically contacted on the front and back sides, whereby typically screen-printed silver and aluminum pastes, respectively, are used. To build the modules, the solar cells are first interconnected by using SnPb solder-coated copper ribbons, and then they are encapsulated between ethylene/vinyl acetate copolymer (EVA) foils. A tempered glass seals the front side, whereas a polymeric back-sheet seals the rear side. Finally, the module is enclosed in an aluminum frame, a junction box is embedded and copper wires are attached [Ziemińska-Stolarska, 2021]. Given the exponential growth in installed PV capacity over the past 20 years, which culminated in the attainment of 2 TW in 2024 worldwide [Philipps, 2024b], it is anticipated that a massive quantity of solar modules will reach the end of their operational lifespan in the forthcoming decades. This scenario has prompted the pursuit of effective methods for recycling PV modules, with the objective of recovering valuable materials.

In this direction, laser sources have been employed to selectively debond targeted materials from the rest of the module. For instance, the glass-EVA front cover could be separated from the silicon wafers by irradiating the module with a ns-pulsed Nd:YAG laser, operating at its fundamental wavelength (1064 nm) or at its harmonics (532 nm or 355 nm) [Bin Anwar, 2024]. As the glass-EVA layer is transparent to those wavelengths, the laser light was absorbed by silicon. The authors of that study reported that the UV wavelength was the most efficient for the debonding process, and explained this observation by the fact that Si presents a short optical penetration depth of only ~10 nm at this UV wavelength, which is orders of magnitude shorter than the penetration depth at the green and infrared wavelengths. Therefore, the controlled energy deposition by the UV laser treatment is localized within the first tens of nm from the Si/EVA interface, allowing for a rapid increase in temperature along with the degradation of the adhesive strength of the EVA layer, but at the same time avoiding thermal decomposition or carbonization of EVA. A similar approach was carried out by Li et al. with the objective to debond the back EVA cover from the Si wafers [Li, 2022b]. In this case, the module was first disassembled in order to remove the Al frame, junction box and back sheet to render the backside EVA accessible. Then, a fiber laser (1064 nm) with a pulse duration of 5 ns was used to irradiate the backside EVA cover. Here, the absorption took place in the Al coating on the backside of the wafer that works as the back electrode. By controlling the fluence dose of the process along with the laser scanning conditions, the EVA could be mechanically peeled without ablating either the Al surface nor compromising the structural integrity of the EVA layer. Moreover, the authors measured the solar cells' performance before and after the dismantling process, showing no device deterioration related to the removal of the EVA. A thorough chemical analysis based on Fourier transform infrared spectroscopy (FTIR), energy dispersive X-ray spectroscopy (EDS), Raman spectroscopy and thermogravimetric (TGA) analysis revealed that the laser treatment had no impact on the microstructure and the bulk properties of the EVA. These findings demonstrate that laser technology can be used to recover not only the EVA layers from the modules, but also the silicon solar cells in operating conditions.

The most valuable material in a commercial mono-Si solar module is arguably the silver used for the electrical contacts, which can amount up to 7.2 g in a typical 250 W module [Jung, 2016]. To recover this precious metal, two laser debonding processes were developed based on two different laser sources, namely a ns-pulsed UV laser and a cw fiber laser operating in the IR (1064 nm). In both cases, the laser beam was scanned over the cell's metal grid, and a spontaneous delamination, i.e., without applying external mechanical forces, of the Ag layer was observed. The study reported that the cw laser operating at 75 W was more efficient than the pulsed UV-laser to remove the silver and also prevented any damage on the underlaying Si surface [Khetri, 2024].

Although thin film technologies accounted for only 2.5% of the worldwide produced PV capacity in 2023 [Philipps, 2024b], the recycling of thin-film solar modules has also acquired a significant importance. While these modules offer significant advantages in terms of reduced material usage and manufacturing flexibility, their composition raises important environmental and resource-related concerns. Many thin-film technologies, such as cadmium telluride (CdTe), copper indium gallium selenide (CIGS), gallium arsenide (GaAs), or hybrid perovskites, rely on elements that are either toxic, such as cadmium, arsenic or lead, or scarce, such as indium and tellurium. These factors underscore the necessity of efficient recycling processes to mitigate environmental risks and recover valuable materials for future use. Laser technology has been proven as a competent pathway to address this challenge by allowing for the selective removal of thin-film layers, and thus the precise separation of toxic or valuable materials from substrates without damaging the underlying components. The removed thin-film materials can be collected and then handled in a separate recycling chain.

# 4. Laser Techniques for Improved Electrochemical Performance

Laser texturing techniques, including Direct Laser Writing (DLW), Direct Laser Interference Patterning (DLIP), and Laser-Induced Periodic Surface Structures (LIPSS), play a significant role in improving the electrochemical performance of materials for energy carrier applications, such as hydrogen evolution reaction (HER). These techniques enlarge surface area and modify wettability, leading to enhanced gas bubble formation and electrochemical behavior. DLW, especially with ultrafast lasers, generates surface patterns with feature sizes ranging from 10 µm to 100 µm as shown in Figure 33.12, enhancing the electrochemical activity of materials like nickel and titanium. Additionally, femtosecond lasers can create superhydrophilic or aerophobic electrodes, increasing the efficiency of HER by improving surface characteristics. DLIP uses laser interference patterning with the advantage of adjustable spatial resolutions and can be applied to three-dimensional (3D) materials such as open-cell metal foams. It has been shown to significantly increase electrochemical capacitance and decrease overpotentials. LIPSS are usually operated at sub-micrometer scale and can further enhance electrode surface area, enabling superior HER performance. Nanostructured surfaces formed through LIPSS exhibit increased capacitance, lower overpotentials, and higher peak current densities in comparison to those without LIPSS-patterned surfaces, thus, hydrogen production can be significantly improved. These laser-based methods can outperform traditional techniques, offering scalable solutions for efficient hydrogen production and other energy carrier applications [Chaudhari, 2017; Gabler, 2018].

In the past decades it has become obvious that global climate change can be attributed by applying new energy sources. Transporting energy globally requires the conversion of energy into energy carriers such as hydrogen, methanol, or others. The process of converting electrical energy into those energy carriers demands increased efficiency and reduced losses in order to be competitive to current fossil fuels such as coal and oil. In that particular field, laser texturing processes show a predominant role for achieving this goal [Chaudhari, 2017].

Especially surface texturing using short- and ultrashort pulsed laser sources was used to carry out a substantial number of research endeavors. In this context, three major laser-based texturing techniques can be identified, which are Direct Laser Writing (DLW, see Chap. 3 (Gräf and Müller)), Direct Laser Interference Patterning (DLIP, see Chap. 9 (Lasagni)) and Laser Induced Periodic Surface Structures (LIPSS, see Chap. 10 (Ancona et al.)), exemplary shown in Figure 33.12a-c. These techniques have been used on two-dimensional (2D) as well as on 3D electrode materials such as flat metal sheets or open cell metal foams [Gabler, 2018; Baumann, 2020]. All those applied techniques depict significant advantages and challenges in comparison to other techniques such as chemical treatment or plasma technologies and are described further in the following sections.

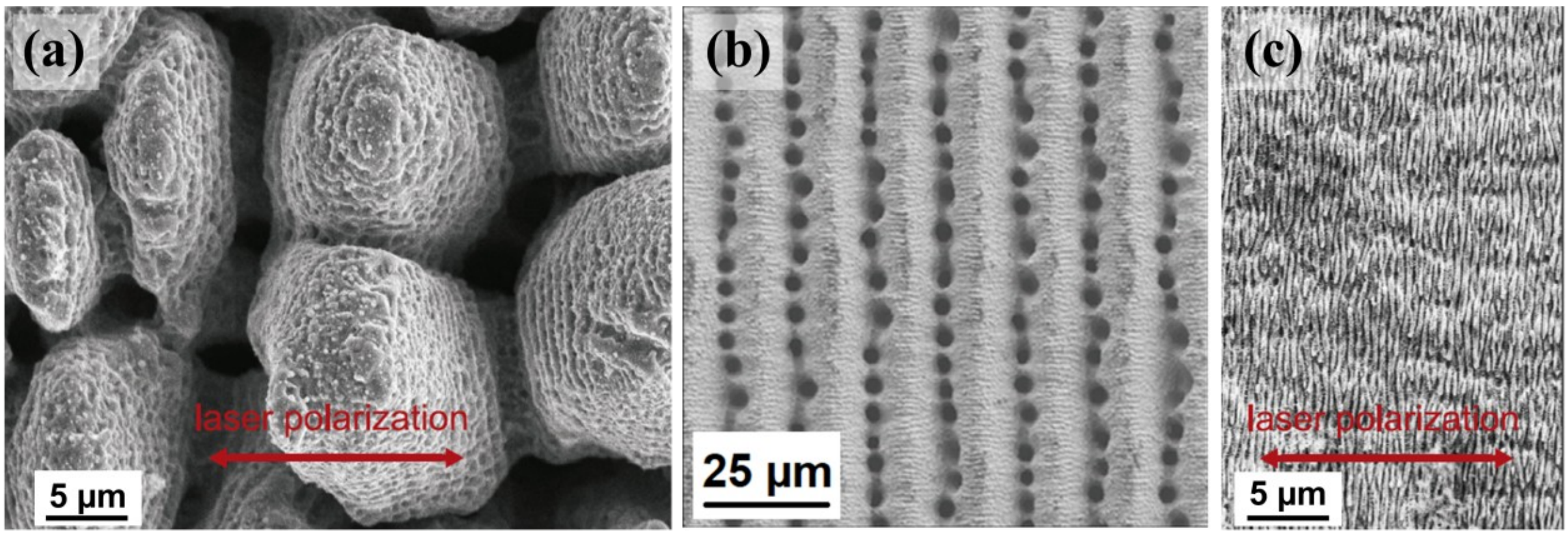

Figure 33.12: SEM Images of central position of Titanium microstructured surface areas on a single spot area with a fluence of F = 0.54 J/cm² and 250 pulses (a) [Gabler 2018], SEM images of DLIP line-like structures with spatial periods (Λ) of 11 µm produced on nickel-foils at a repetition rate frep = 8.5 MHz and scanning speed vscan = 200 m/s and 250 pulses (b) [Raenke 2024a], SEM Images of outer position of Titanium nanostructured surface areas on a single spot area with a fluence of F = 0.54 J/cm² and 250 pulses (c) [Gabler 2018] (a, c) Reprinted from [Gabler, 2018], Gabler A., Müller C.I., Rauscher T., Gimpel T., Hahn R., Köhring M., Kieback B., Röntzsch L., Schade W.: Ultrashort-pulse laser structured titanium surfaces with sputter-coated platinum catalyst as hydrogen evolution electrodes for alkaline water electrolysis, International Journal of Hydrogen Energy 43, 15, 10826-10833, (2018), Copyright 2018, with permission from Elsevier. (b) Reproduced from [Raenke 2024a], Ränke F., Baumann R., Voisiat B., Soldera M., Lasagni A.F.: Nano/microstructuring of Nickel Electrodes by Combining Direct Laser Interference Patterning and Polygon Scanner Processing for Efficient Hydrogen Production, Advanced Engineering Materials 26, 10, 2301583, (2024), with permission from Wiley, Copyright 2024.

## 4.1 Coarse Textures for Electrochemical Improvements by Direct Laser Writing

Nowadays, the most prominent power to gas (PtG) application, due the political ambitions to tackle the global rising need of carbon neutral energy, is the hydrogen evolution reaction or "HER". For improving this electrochemical process, different laser-based texturing approaches has been endeavored. The usage of DLW leads to the generation of surface patterns with feature sizes between 10 and 100 µm. As was reported by several research groups, laser beams were employed on chemical active materials such as nickel (Ni), titanium (Ti), or alloys composing molybdian-sulfur ($MoS_2$), cobalt-nickel (Cr-Ni), or iron-nickel for producing different geometries [Faubert, 2015; Gabler, 2018; Wang, 2024; Raenke, 2024a]. Laser-based textures lead to topographical increase of the surface area, resulting in improved electrochemical behavior such as higher capacity or lower overpotentials and change of wettability which leads to different gas bubble formations. The most prominent laser surface texturing technique is DLW. Here, in most cases a single laser beam is guided over the surface area of electrodes by means of a X-Y-axis system or highly dynamic galvo- or polygon scanners. In addition, ultra-fast laser sources in the pico, sub-pico and femtosecond range were applied to generate various types of surface architectures for increasing the chemically active area of an electrode. However, due to the high thermal diffusion length as well as enlarged melt formations, nanosecond lasers were rarely employed for this application. As shown in Figure 33.13, the spatial period of the DLW generated structures was not possibly below 10 µm due to the limitation of laser beam focusing. Furthermore, the interaction of the processed material with the laser radiation also leads to different surface conditions. For example, ultrashort pulses of 400 fs and 350 fs were applied on the surface texturing of MoS2 and Cr-Ni, as illustrated in Figures 33.13a,b and Figure 33.13c,d, respectively. The appearance and surface quality are important factors in terms of gas evolution reactions, since it defines the amount of gas bubble nucleation sites. Ando et al. [Ando, 2022] reported that a laser-based surface texture leads to different gas bubble formations depending on the location of the laser textured area.

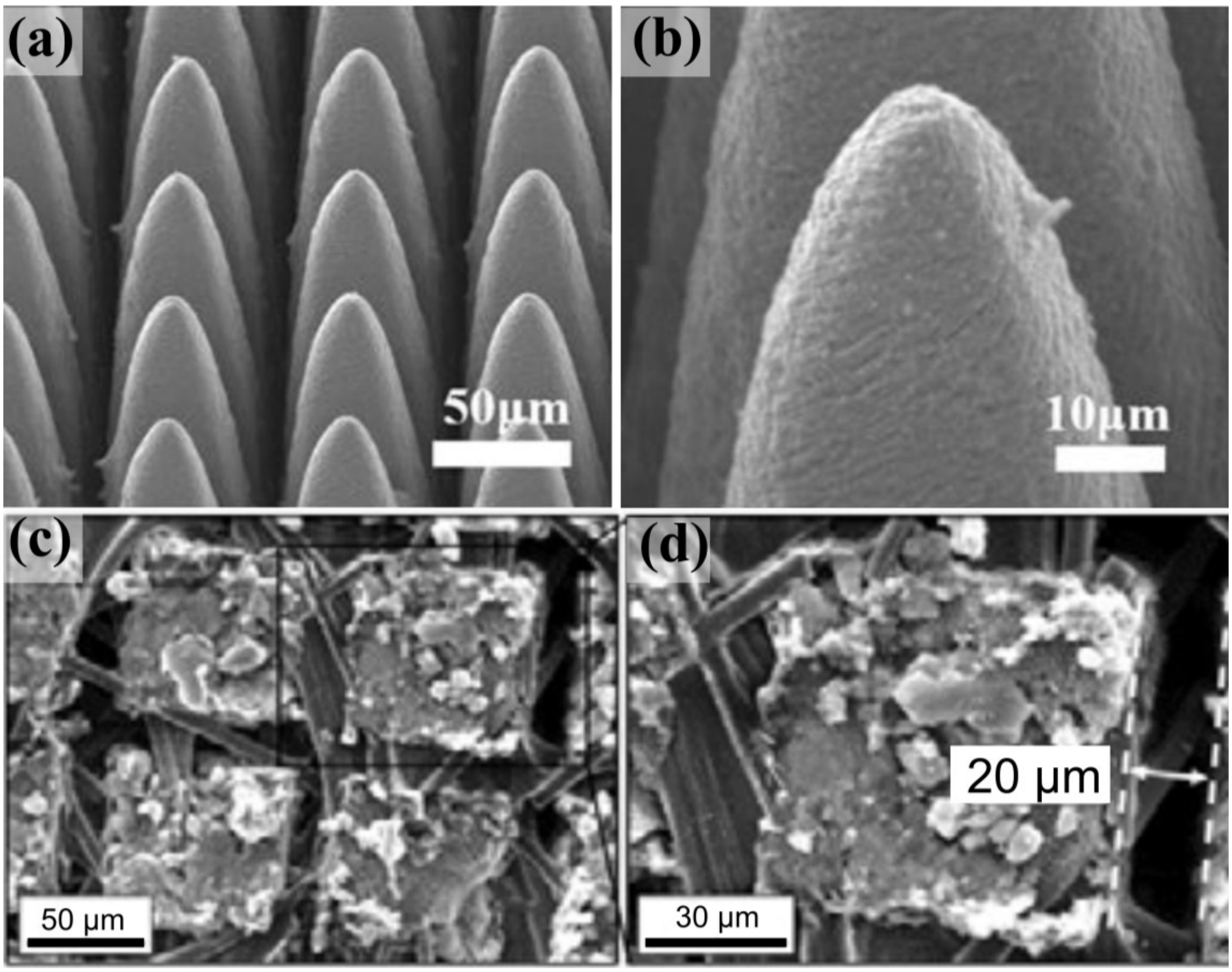

Figure 33.13: SEM images of (a) femtosecond structured-Molybdian (Mo) and $MoS_2$/fs-Mo (b) [Wang 2024] (b), and top-view SEM image of an ultrafast laser structured 3D micro-grids in Cr-Ni electrode (c,d), [Faubert, 2015]. (a, b) Reprinted from [Wang, 2024], Wang Z., Song L., Tao H., He Y., Yang Y., Wang T., Yu H., Lin J., Dong X.: Fabrication of micro/nano wettability $MoS_2$ hydrogen evolution electrocatalyst based on femtosecond laser, International Journal of Hydrogen Energy 51, 327-335, (2024) Copyright 2024, with permission from Elsevier. (c, d) Reproduced from [Faubert, 2015] Faubert P., Müller C., Reinecke H., Smyrek P., Proell J.,Pfleging W.: Femtosecond laser structuring of novel electrodes for 3D fuel cell design with increased reaction surface, Mater. Res. Soc. Symp. Proc. 1777, 7-13, (2015), with permission of Springer Nature. Copyright 2015.

Highly efficient and durable superhydrophilic and superaerophobic electrodes were synthesized by growing $MoS_2$ catalysts on molybdenum (Mo) substrates textured with femtosecond laser. The application using femtosecond laser processing increased the number of active sites, enlarged the surface area, and thus enhanced the hydrogen evolution efficiency [Wang, 2024]. This technique also imparted superhydrophilicity and superaerophobicity to the electrodes, facilitating rapid bubble detachment and efficient water replenishment. The use of laser with a pulse duration of 350 fs generated rectangular box-like structures, as shown in Figure 33.13c, d. A large amount of recast material formation was observed in the processing zone adjacent to the laser-induced trenches. Compared to unstructured electrodes, the structured surfaces exhibited a 30% increase in surface area, which resulted in a corresponding 30% reduction in ohmic resistance, thereby enhancing the efficiency of the electrochemical reaction. Further investigation is required to determine whether the recast material contributes positively or negatively to the overall process [Faubert, 2015].

## 4.2 Impact of Micro- and Sub-Micrometer Structures onto the Electrochemical Performance of Electrodes

In addition to DLW, a similarly versatile processing technique for varying the spatial resolution of electrode material texturing is DLIP. As previously described, by controlling the interference angle and polarization of two or more laser beams, unique periodic surface patterns can be generated. These patterns increase the surface area and enable tailored wetting properties that influence bubble generation and dynamics. Similar to DLW, DLIP typically employs ultrafast laser sources with pulse duration in picosecond and femtosecond range for the applications in gas evolution reactions. Historically, this technique was primarily applied to flat materials due to the limitations of laser beam superposition in a 2D plane [Eckhardt, 2013]. Afterwards, this technology was first introduced for treating open-cell metal foams to enhance electrochemical behavior, as shown in Figure 33.14.

Exemplary SEM images of a DLIP-treated metal foam (processed with 98.6% overlap, 3.3 J/cm² laser fluence, and a spatial period of 5.8 µm) demonstrate the successful application of this method to complex geometries, shown in Figure 33.14 a,b. In addition to the DLIP-generated fringes, supplementary surface patterns known as LIPSS were observed. These structures form when ultrashort-pulsed laser radiation interacts with metallic surfaces [Vorobyev, 2013; Waldvogel, 2018]. Consequently, the LIPSS features, alongside the DLIP fringes, create a periodic nano-roughness that further increase the overall surface area.

Two types of LIPSS features are generally identified: low-spatial-frequency LIPSS (LSFL) and high-spatial-frequency LIPSS (HSFL). The LSFL structures with a spatial size of approximately 810 nm correspond closely to the applied laser wavelength, which is typical for these structures. Conversely, the HSFL structures oriented perpendicular to the DLIP features exhibit a spatial period of about 130 nm, much smaller than the laser wavelength, which is characteristic of such patterns on metallic surfaces [Aguilar, 2018]. The cyclic voltammograms recorded at varying scan rates for unstructured and DLIP-structured foams are illustrated in Figures 33.14c. For both types of electrodes, nearly rectangular cyclic voltammograms were obtained, indicating the absence of significant Faradaic reactions and a suitable potential window for determining the double-layer capacitance. The double-layer capacitance ($C_{dl}$ [µF cm$^{-2}$]) was calculated by plotting the average of the absolute anodic ($j_a$) and cathodic ($j_c$) current densities against the scan rate ($v$ [V s$^{-1}$]), as shown in Figure 33.14c, using the following equation [Gileadi, 2011]:

$$C_{dl} = \frac{j_a + |j_c|}{2 \cdot v} \tag{33.1}$$

The unstructured foam exhibited a capacitance of approximately 702 µF/cm², whereas the DLIP-structured foam demonstrated a capacitance of 1330 µF/cm², nearly double than that of the unstructured foam [Baumann, 2020]. Electrochemical impedance spectroscopy (EIS) was employed as a complementary method to measure the double-layer capacitance. Figure 33.14d presents the Nyquist plots of the impedance data for both unstructured and structured foams. In both cases, a depressed semicircle was observed, which is attributed to the porosity or surface roughness of the electrode [Orazem, 2008]. The data were approximated using a modified Randles circuit (Figure 33.14d), where the capacitance was replaced with a constant phase element (CPE). The double-layer capacitance was calculated using equation 33.1 [Orazem, 2008]. It is worth noticing that here only the top surface layer of the 3D foam was textured due to the size of the interference volume.

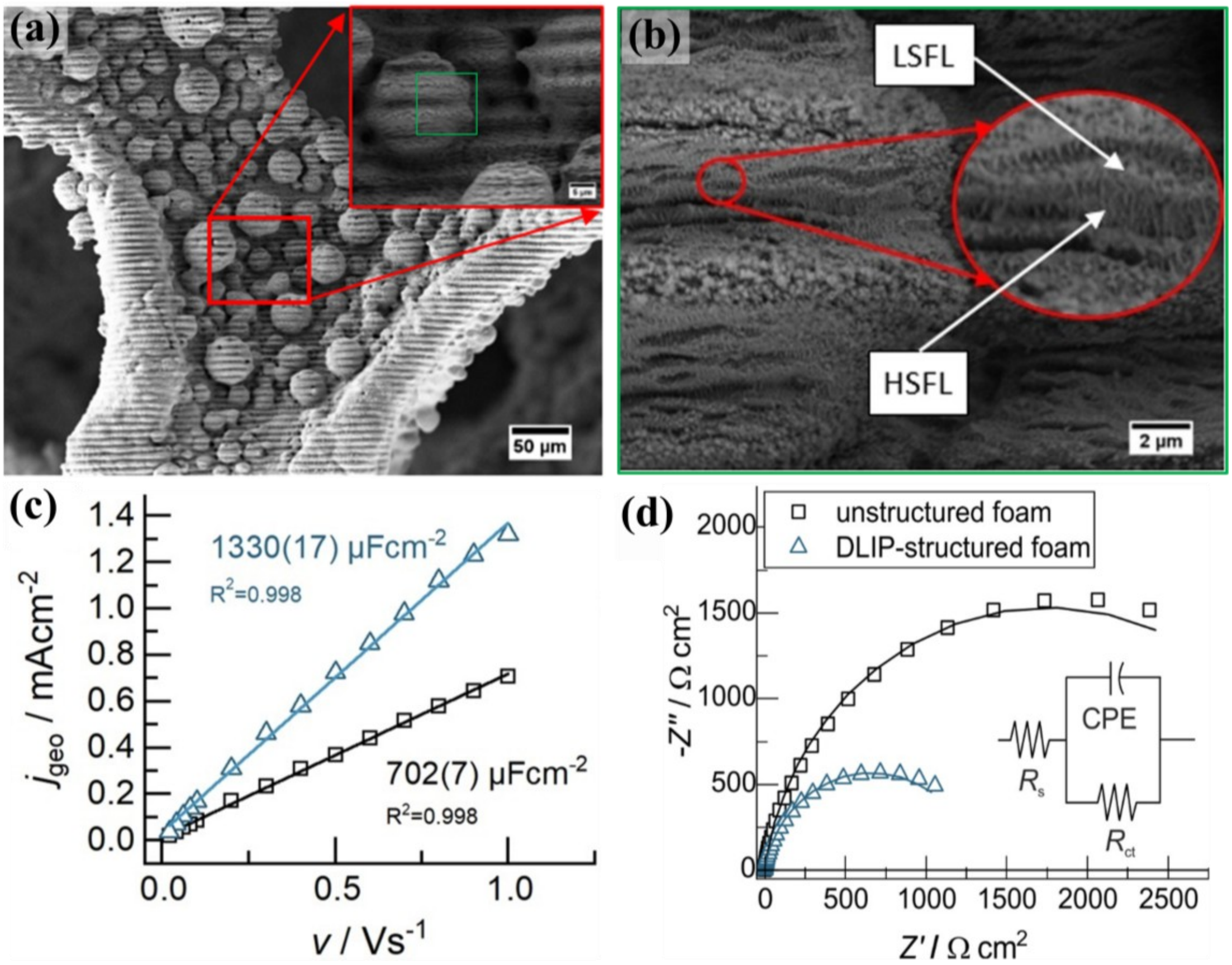

Figure 33.14: SEM images of open-cell foam out of steel with line-like pattern produced using DLIP (a). The insets in (a) show a magnification of the highlighted area for better visualization; magnification of the green highlighted section of the inset for better visualization of high (HSFL) and low (LSFL) spatial frequency LIPSS (b); overlap: 98.6%; fluence: 3.34 J/cm²; spatial period: 5.8 µm, The average of the absolute anodic and cathodic current densities versus the scan rate (c), Complex plane plot (Nyquist plot) for unstructured and DLIP structured foam in a $N_2$-purged 29.9 mass-% KOH solution at 60°C. Symbols are experimental points and solid lines are fits by the Randles CPE (Constant Phase Element) model (d). [Baumann, 2020] (a,b,c,d) Reproduced from [Baumann, 2020] with permission from Japan Laser Processing Society.

Recent developments in processing heads however eliminated this drawback of the limiting interference volume, giving the opportunity to texture 3D materials without the adjustment of the focus plane. Line- and pillar-like DLIP textures with a spatial period of ~3.0 µm were firstly fabricated on the Ni-foils, as shown in Figure 33.15a and b. The pulse duration τ [s] and the number of consecutive passes N [-] were varied to evaluate their influence on the structure morphology.

As previously mentioned, the laser repetition rate $f_{rep}$ [Hz] was maintained constant throughout all experiments. During the investigations, the pulse energy was gradually increased from 41.1 µJ to 90.8 µJ. The total energy applied per unit area, referred to as the cumulative laser fluence ($\Phi_{cum}$ [J/cm²]), was calculated. The measured structure depths for treated samples are shown in Figure 33.15c and d. The depth of the line-like DLIP textures was observed to increase almost linearly with the cumulative fluence up to ~8 J/cm², and this trend was consistent across all pulse durations.

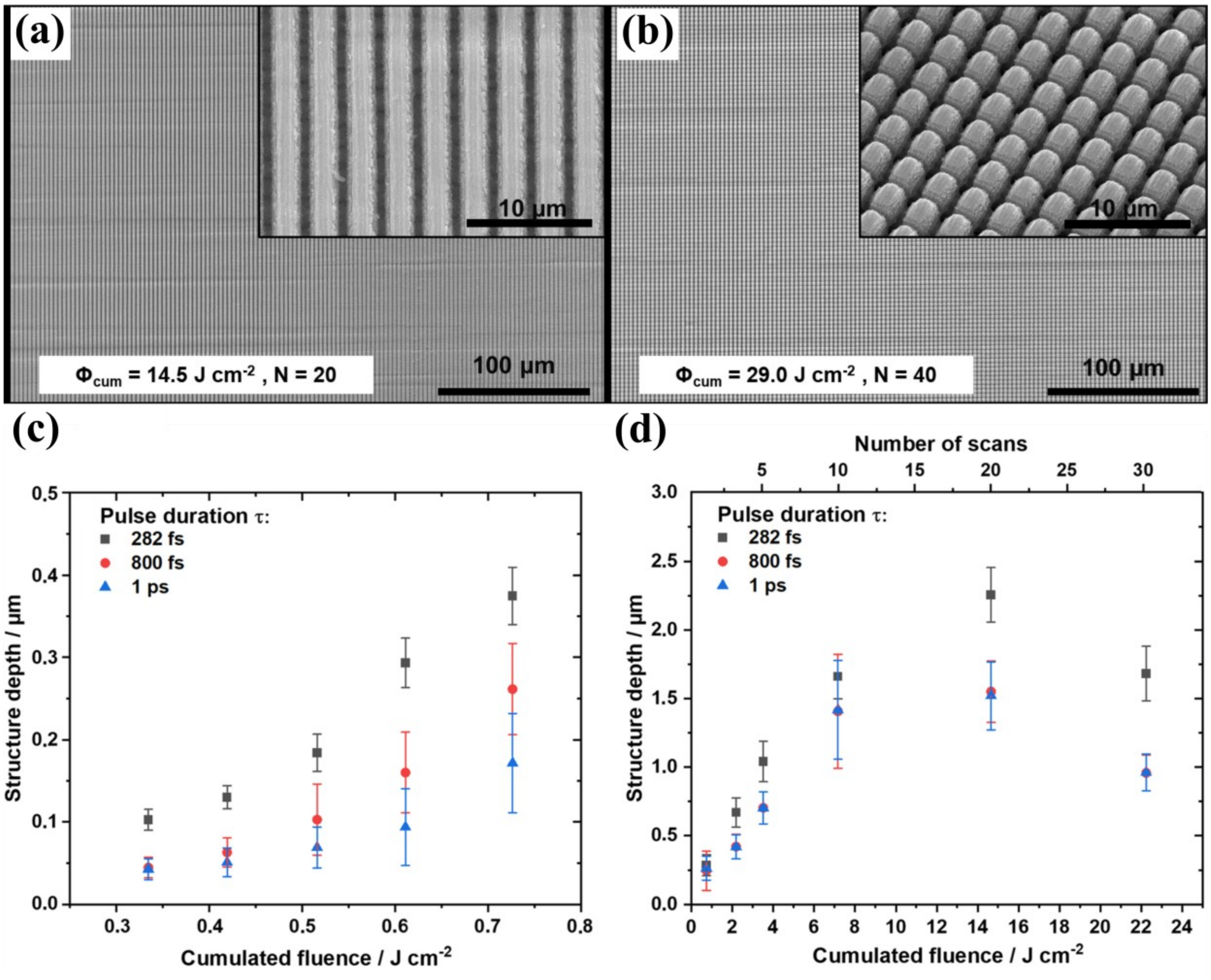

Figure 33.15: SEM images of DLIP line-like (a) and cross-like (b) structures produced on Ni-foils at a pulse duration τ = 282 fs, repetition rate $f_{rep}$ = 1 kHz, overlap (OV) = 95 % and hatch distance (HD) = 139 µm. The cumulated fluence ($\Phi_{cum}$) and the total number of applied scans (N) are displayed in the corresponding labels, Structure depth as function of the cumulated fluence for single scan (c) and multiple scans (d), using pulse durations of 282 fs, 1 ps and 10 ps [Raenke, 2024b]. Reproduced from [Raenke, 2024b] with permission from Japan Laser Processing Society.

Besides, the deepest DLIP features were generated over the entire range of cumulative fluence values at the shortest pulse duration of 282 fs. For example, at the maximum fluence of 0.73 J/cm², a structure depth of 0.37 ± 0.03 µm was achieved with a pulse duration of 282 fs. In comparison, a structure depth of 0.26 ± 0.06 µm and 0.17 ± 0.06 µm were observed for an 800 fs and a 1 ps pulse duration, respectively. These results indicate that the structure depth at 282 fs was approximately twice than that achieved at 1 ps. It is well-established that shorter laser pulses, particularly those with durations below 1 ps, require less energy per unit area to initiate ablation at the material surface. This is attributed to the efficient utilization of absorbed energy for the ablation process, with minimal thermalization of the remaining material [Stuart, 1996]. Consequently, shorter pulse durations enhance the ablation rate, as has been reported for various metals [Winter, 2021]. When the samples were scanned multiple times, deeper structures were produced due to increased material removal at the interference maxima positions. However, when the number of passes was increased to 30, the structure depth decreased regardless of the pulse duration. This behavior has been reported in DLIP structuring involving nanosecond pulses, nontheless, the underlying mechanisms for this reduction in structure depth remain unclear [Raenke, 2024b].

Ni-electrodes with line and cross-like structures were fabricated and evaluated for HER [Raenke, 2024b]. The obtained results indicated that the DLIP-treated electrodes benefited from enlarged electrochemically active surface area and enhanced intrinsic activity which led to a considerable decrease of the overpotential of up to 51% compared to the untreated Ni-electrode. Furthermore, an increase in the double-layer capacitance up to ten times was also observed. These findings indicate that DLIP generated with femtosecond laser source can be employed to fabricate Ni-electrodes at industrial scale for highly efficient production of hydrogen.

## 4.3 Nanotextured Surfaces to Improve Electrochemical Behavior

The smallest laser generated periodic surface patterns that can be reached are laser-induced periodic surface structures, so called "LIPSS". The dimensions of these sub-micro or nanometer scale architectures is mainly depend on the applied laser wavelength. That means that the flexibility of changing the spatial period of LIPSS mainly limited by the laser beam source. Furthermore, LIPSS are generated when the applied laser fluence is just below the ablation threshold fluence of the radiated material. For improving the electrochemical performance of gas evolution reaction in terms of Hydrogen Evolution Reaction (HER) and Oxygen Evolution Reaction (OER), LIPSS were applied by several research groups. Some results are shown in Figure 33.16.

In the study of Ahmad et al., Ni sheets were subjected to irradiation using ultrashort laser pulses, resulting in the formation of LIPSS, as observed using SEM (Figure 33.16 a-f). The nanostructured surfaces exhibited a periodicity of 570 ± 30 nm, with the presence of nanoparticles (protrusions) on top, which were significantly smaller in size [Ahmad, 2024]. The electrochemical performance of these laser-treated Ni electrodes was then evaluated and notable improvements were observed. Specifically, the electrodes with nanostructured surfaces exhibited a 5.3-fold increase in double-layer capacitance, lower overpotentials at high current densities, and higher peak current densities in cyclic voltammetry analyses. Moreover, the electrodes with LIPSS produced 3.7 times more hydrogen gas compared to untreated samples, and the enhanced hydrogen production rate was attributed to the increase in effective electrochemical active surface area of the nanostructured electrodes. These improvements were further characterized by steadier and higher current densities during the HER [Poimenidis, 2021]. The laser-induced textures shown in Figure 33.16 also contribute to the enhancement in electrochemical capacity of these alloys. The laser surface structuring process served two key purposes: (i) enriching the oxidation states of the material through tailored laser-matter interactions, and (ii) increasing the electrochemically active surface area by inducing periodic nanostructures on the surface. The results indicated that the multi-principal element alloys, benefiting from the "cocktail effect" in oxidation states, exhibited a significant increase in electrochemical activity for water splitting, as well as enhanced corrosion resistance and mechanical strength. Notably, the electrochemical performance of these alloys rivaled that of expensive platinum/graphite (Pt/C) and iridium/graphite (Ir/C) electrodes, with a measured potential of 1.61 V at a current density of 10 mA/cm² for the laser-structured NiCo(Cr/V) alloy, compared to 1.58 V at the same current density for the Pt/C and Ir/C bifunctional electrodes. The intrinsic diversity in oxidation states of these alloys was further improved by femtosecond laser surface structuring, which not only increased the electrochemical surface area but also enhanced key electrochemical parameters including reduced overpotentials and Tafel slopes [Ahmad, 2024].

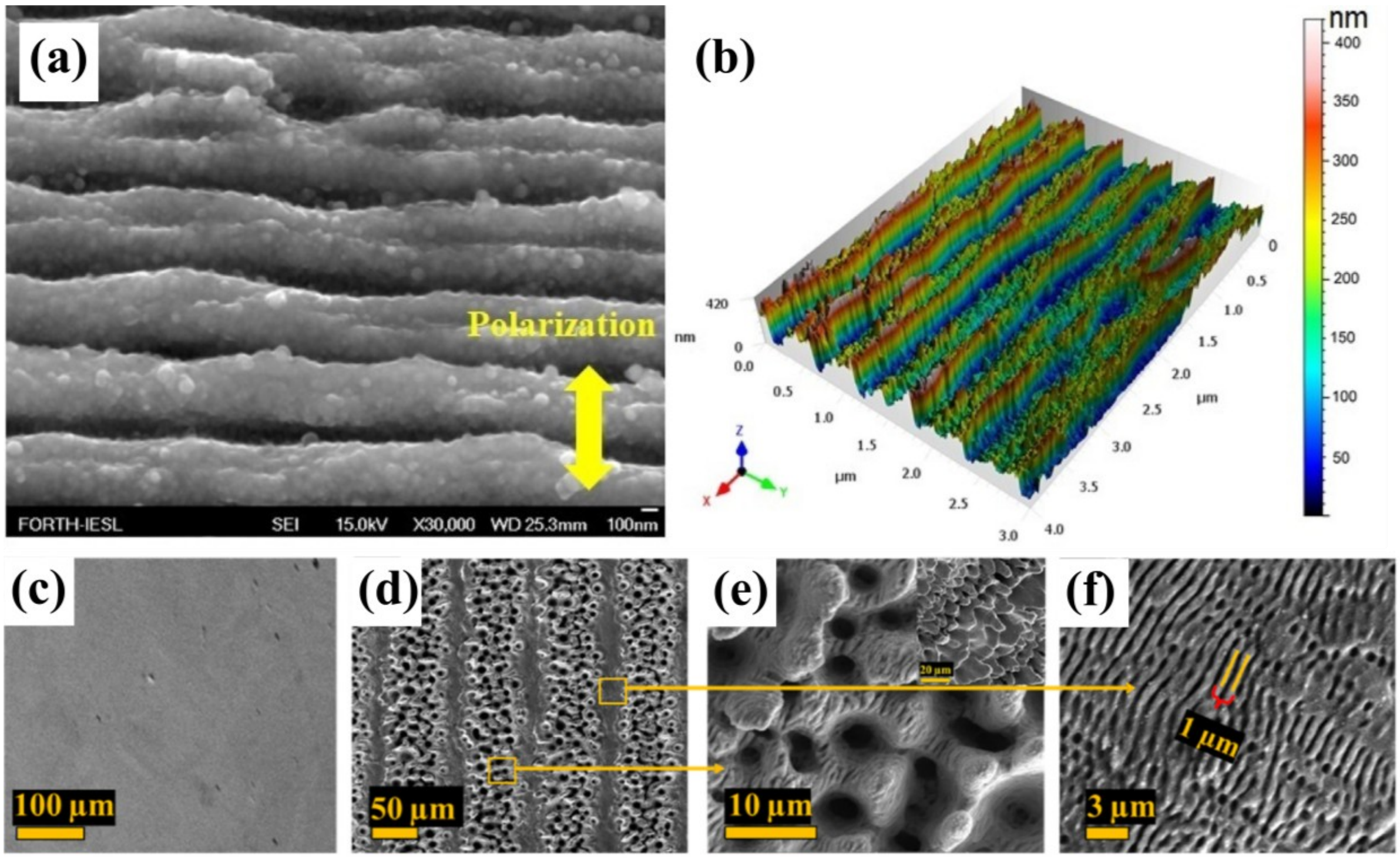

Figure 33.16: SEM images for the laser-structured electrode with 10°angle tilted (a), 3D interactive topography image of the nanostructured electrode (b) [Poimenidis, 2021], SEM analysis of pristine and LS-MEA, NiCo(Cr/V): Polished surface (c), fs-laser-structured electrodes showing two distinct regions (d): The highest energy region (gaussian profile) (e), and the LIPSS-dominated region due to the overlap of low energy spots of the laser beams in raster scanning rough areas with microgrooves (f) [Ahmad, 2024] (a, b) Reprinted from [Poimenidis, 2021] Poimenidis I.A., Tsanakas M.D., Papakosta N., Klini A., Farsari M., Moustaizis S.D., Loukakos P.A.: Enhanced hydrogen production through alkaline electrolysis using laser-nanostructured nickel electrodes, International Journal of Hydrogen Energy 46, 37162-37173, (2021), Copyright 2021, with permission from Elsevier. (c, d, e, f) Reprinted [Ahmad, 2024] Ahmad S., Boltaev G., Egilmez M., Abuzaid W., Alawadhi H., Kannan A.M., Alnaser A.S.: Impact of femtosecond laser surface structuring on NiCoCr and NiCoV medium entropy alloy systems for an overall electrochemical water splitting, International Journal of Hydrogen Energy 59, 1094-1105, (2024), Copyright 2024, with permission from Elsevier.

# 5 Laser Technology for Advanced Battery Systems

Like no other, lithium-ion batteries are currently in the focus of social development, politics and research, mainly driven by electromobility and stationary storage solutions for sustainable energy generation. This section shows how advanced laser technology can provide groundbreaking improvements in battery manufacturing, not only to make production more sustainable, but also to significantly improve battery properties and make a significant contribution to increased safety, sustainability and cost-effectiveness across the process chain. In particular, ultrafast laser structuring is reviewed and evaluated with regard to current issues and scalable solutions for future industrial production.

## 5.1 Objectives for Next Generation Lithium-Ion Batteries

To meet the ambitious goal of achieving zero emissions by 2050 in Europe, the shift from fossil fuels to renewable energy is inevitable [Fetting, 2020]. The transition to sustainable energy sources with enhanced energy efficiency play a key role in this revolution. One of the most important measures to drive this transformation is the electrification of the transport and energy sectors, in which batteries play a pivotal role. Among various types of batteries, rechargeable lithium-ion batteries (LIBs) have garnered significant attention and are widely applied to power stationary grids, portable devices, and electric vehicles [Armand, 2020]. The production capacity of battery cells in Europe is expected to increase from 140 GWh in 2023 to about 1 TWh in 2030 as demand continues to grow [Bielewski, 2022]. Worldwide, the lithium-ion battery market will grow from 250 GWh in 2020 and 1,200 GWh in 2023 to 3,200 GWh in 2030, with a Compound Annual Growth Rate $CAGR_{20\text{-}30}$ of 26% in volume [Wolf, 2024]. Besides, fast charging and a long lifetime should also be guaranteed, which means that the cells should reach 80% of their State-Of-Charge (SOC) in 10 min, while achieving about 2000 cycles under high operating voltage up to 4.8 V by 2030. In addition, the cost per battery pack in Europe need to be reduced to below 75 €/kWh by 2030 while enhancing the sustainability of raw material processing and advanced materials manufacturing [Bielewski, 2022].

A central question in this context is how laser technology can contribute to accelerating the transition to clean energy through advances in lithium-ion battery production. The answer is multifaceted. On the one hand, laser processing has already been introduced into LIB manufacturing, for example for laser welding of bus bars and laser cutting of electrodes [Feng, 2023; Keppeler, 2021]. On the other hand, laser materials processing has a huge potential to provide additional and performance-enhancing new technological solutions in battery production, providing essential support to achieve the above goals. For example, laser drying of wet-coated electrodes, which is at an early stage of development, is in principle capable of significantly reducing the energy consumption for electrode drying by more than 50%, making the entire production process more sustainable [Neb, 2022]. However, a real boost in electrochemical performance for the next generation of batteries will also require innovative battery designs such as the 3D battery and the thick film concept, which for the first time enable the production of batteries that can be used for both high energy and high power applications, while reducing the amount of inactive material and increasing cell lifetime and safety [Pfleging, 2020; Sterzl, 2024].

## 5.2 Principle and Components of Lithium-Ion Batteries

Lithium-ion batteries (LIBs) are electrochemical energy storage systems composed of multiple individual secondary cells connected in parallel and/or series. These systems enable the reversible storage of chemical energy, which can be converted into electrical energy. The nearly complete reversibility in these rechargeable cells is enabled by the rocking-chair mechanism, based on the diffusion of lithium ions between the anode and cathode during charge and discharge cycles. Despite significant advancements in electrode materials and cell designs, the fundamental components of LIBs have remained largely unchanged. An typical lithium-ion cell consists of two electrodes (anode and cathode), current collectors (copper foil for the anode and aluminium for the cathode), a separator, and an electrolyte, as illustrated in 33.17. The electrodes and separator are immersed in a liquid electrolyte. In solid-state batteries, however, the separator and liquid electrolyte are replaced by a solid-state electrolyte. All cell components are enclosed within an insulated housing, which isolates them from the external environment.

During discharging, lithium ions migrate from the anode's active material through the electrolyte, cross the ionic permeable separator, and intercalate into the cathode's active material. Simultaneously, electrons flow from anode to cathode via an external circuit. This process involves reduction at the cathode and oxidation at the anode, whereby the reactions are reversed during the charging process. Electrodes have porous composite structures composed primarily of active materials, conductive agents, and binders. Currently, graphite is the predominant anode material, while $LiFePO_4$ (LFP) or $LiNi_{0.8}Mn_{0.1}Co_{0.1}O_2$ (NMC 811) are widely used cathode materials. These active materials typically account for over 90 wt.% and up to 96 wt.% of the electrode and play a key role in determining the energy density of the cell. Conductive agents, commonly referred to as carbon black or conductive carbon, are typically applied in electrodes at concentrations below 5 wt.% to reduce the electrical resistivity of composite electrodes [Spahr, 2011].

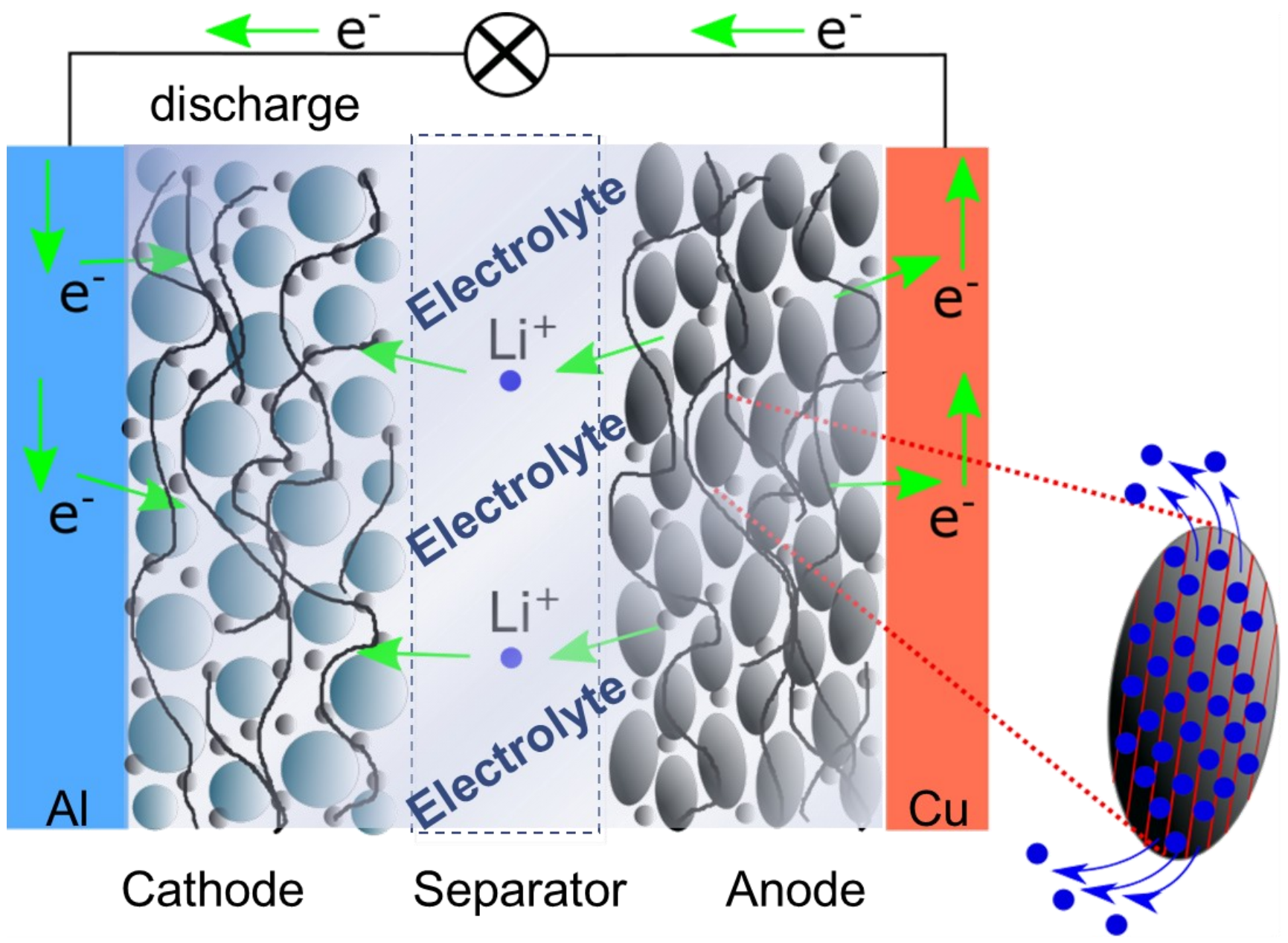

Figure 33.17: Schematic representation of a lithium-ion cell during discharge. The cathode contains active material consisting of spherical ceramic particles, while the anode is represented by flattened or flaky graphite particles. The enlargement shows a single graphite particle with its basal surfaces and the lithium-ions stored within. © KIT

Binders play a critical role in holding active materials and conductive agents together with the current collector, thereby ensuring the mechanical stability of electrodes. By combining conductive agents with binders, conductive pathways can be established, preventing the isolation of active material particles [Bauer, 2015]. Although binders constitute normally below 5 wt.% in commercial electrodes, they are essential for performance metrics such as battery lifetime and rate capability [Chou, 2014]. For cathodes, the battery industry predominantly uses polyvinylidene fluoride (PVDF) as binder, while water-soluble binders like sodium carboxymethyl cellulose (CMC) and styrene-butadiene rubber (SBR) are widely employed for anodes.

The liquid electrolyte in LIBs typically consists of a conductive salt, such as lithium hexafluorophosphate ($LiPF_6$), dissolved in a mixture of organic solvents including ethylene carbonate (EC), ethyl methyl carbonate (EMC), diethyl carbonate (DEC), and dimethyl carbonate (DMC) [Armand, 2020]. Additionally, modern electrolytes incorporate additives such as fluoroethylene carbonate (FEC) [Liu, 2019] or vinylene carbonate (VC) [Xia, 2015], which facilitate the formation of a stable and homogeneous solid electrolyte interphase (SEI) on anodes [Zhang, 2006]. Further enhancements include the addition of flame-retardant additives to mitigate self-heating reactions triggered by electrolyte decomposition, thereby reducing the risk of thermal runaway [Dagger, 2017].

Separators are porous membranes that serve as physical barriers between cathodes and anodes. They have typical thicknesses ranging from 10 to 30 μm [Deimede, 2015]. Their primary function is to prevent short circuits within the cell. To ensure reliable performance, separators must exhibit chemical and electrochemical stability, remaining inert under the strong oxidizing and reducing conditions encountered during charging and discharging, without generating impurities. Separators are generally categorized as monolayer, multilayer, and modified separators (e.g., with ceramic coating or surface modification). Polymers such as polypropylene (PP) and polyethylene (PE) are commonly used as the basic framework of separators [Venugopal, 1999].

# 5.3 Active Materials for Next Generation Lithium-Ion Batteries

The olivine-type material $LiMn_xFe_{1-x}PO_4$ (LMFP) and spinel-type material $LiNi_{0.5}Mn_{1.5}O_4$ (LNMO) have recently received significant attention due to their high working voltage of 4.1 V and 4.7 V, respectively, which results in enhanced power densities [Yang, 2021, Liang, 2020]. Although challenges like rapid capacity fading and electrolyte decomposition at high operating voltages remain, these materials show great potential as next-generation high-energy cathode materials.

As for anodes, graphite with either spherical or flake-like particles remains the most widely used commercial active material for LIBs. The theoretical specific capacity of lithium-intercalated graphite $LiC_6$ is 372 mAh/g [Asenbauer, 2020]. However, silicon (Si)-based materials such as Si-added graphite composites, pure Si nanoparticles or $SiO_x$ nanoparticles, are gaining more attention due to the high theoretical capacity of Si exceeding 3500 mAh/g [Li, 2021]. The energy density of LIBs can be significantly enhanced by applying Si-containing anode materials. Recent developments have focused on integrating silicon into graphite anodes to further improve performance [Zheng, 2019, Meyer, 2021].

To meet future targets, Si content in anodes should exceed 20 wt.% by 2030, enabling specific capacities above 1000 mAh/g [Bielewski, 2022]. However, substantial technological challenges remain, particularly the high-volume expansion of Si during lithiation, fast charging ability, and the stability of SEI during electrochemical cycling.

## 5.4 State-of-the-Art of Battery Production

Battery production involves multiple sequential fabrication steps, covering production lines of electrode manufacturing, cell assembly, and cell priming. Figure  shows these typical manufacturing steps for a lithium-ion battery with pouch cell design. For cells with other designs such as coin cells, cylindrical cells, or prismatic cells, the cell assembly procedure as well as cell priming can be slightly different.

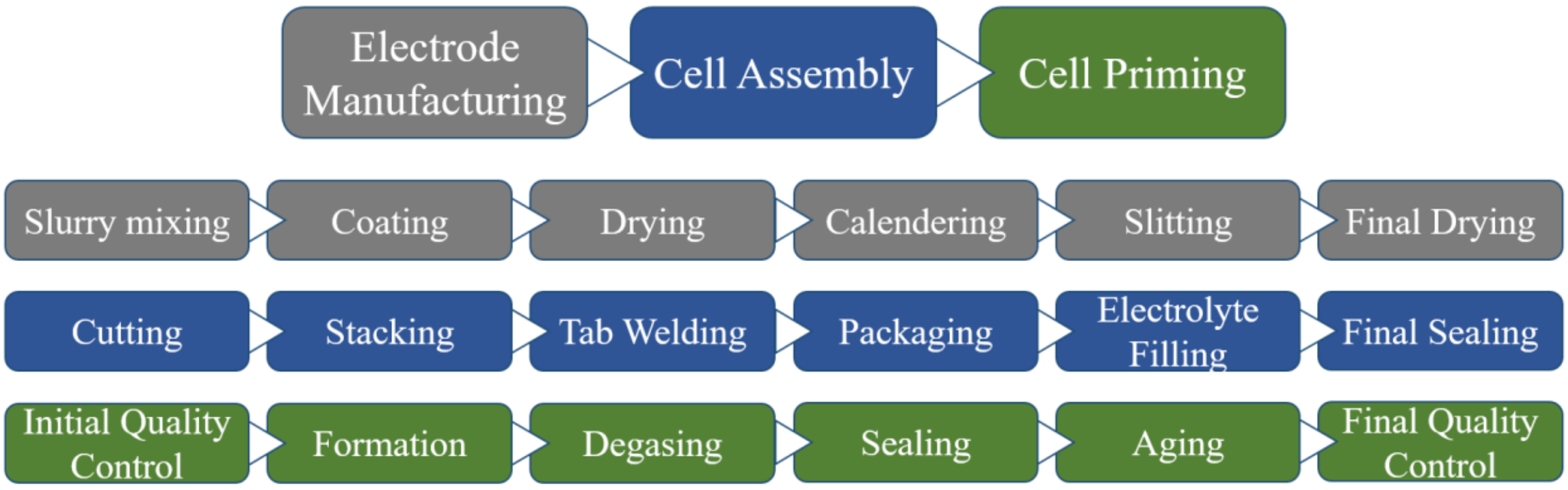

Figure 33.18: Battery manufacturing chain for LIBs from slurry mixing to cell priming

It is worth mentioning that the traditional manufacturing concept offers several integration points for laser processing, such as laser drying after coating, laser cutting, slitting, or notching of electrodes, and laser structuring either before or after electrode calendering. During the calendering process, dried electrodes are passed between rolls under applied pressure and temperature. As a result, electrode thickness and porosity are reduced. This step is essential for enhancing volumetric energy density and increasing the electronic conductivity of the electrodes. Calendering also influences the wetting behavior of the liquid electrolyte in composite electrode. For graphite anodes, moderate calendering improves the wetting behavior due to preferential particle alignment [Sheng, 2014]. However, excessive calendering can reduce the wetting rates for both graphite anodes and cathodes, likely caused by the closure of micropores within the electrodes [Davoodabadi, 2019].

For cell assembly the electrodes are cut into appropriate dimension and are stacked afterwards with the separators placed between cathodes and anodes. Various winding, stacking, and folding techniques can be applied for the assembly of pouch, prismatic, and cylindrical cells. The number of electrodes is adjusted based on the desired capacity at the cell level. In contrast, coin cells typically consist of a single cathode and a single anode. For cells with multiple layers of electrodes, the anodes and cathodes are welded to their respective terminals using ultrasonic or laser welding [Lee, 2010].

The stack is then enclosed within a housing such as a pouch bag or hard case to isolate them from the external environment, followed by the injection of electrolyte. The amount of electrolyte is usually determined by multiplying the pore volume of separators and electrodes by a factor of at least 1.4, to ensure a complete SEI formation and an optimal ionic conductivity. Insufficient electrolyte leads to capacity loss and reduced lifetime, while excess electrolyte lowers the energy density at the cell level [Günter, 2019].

Once the housing is sealed, the cells are stored under controlled temperature to enhance electrolyte wetting and facilitate gas diffusion [Knoche, 2016]. Subsequently, the formation process begins typically with a low charging current. During this process, a stable SEI layer forms on the anode. This is followed by an aging phase lasting two to three weeks. In addition, quality control measures are conducted, including inspection of cell voltage, capacity, and impedance before, during, and after cell priming, to identify and exclude cells with short circuits or insufficient capacity [Kwade, 2018].

# 5.5 Laser Processes in Battery Manufacturing

Lasers are employed for a wide range of applications in battery manufacturing processes, including welding, cutting, slitting, notching, drying, annealing, structuring and surface modification (see Figure 33.19). Laser welding has already been implemented in battery industry for joining copper and aluminum tabs with multiple current collector flags and for sealing battery housings [Sakagawa, 2011, Schmidt, 2016]. Compared to conventional ultrasonic welding, laser welding offers contactless, rapid processing with minimal localized heating, making it an efficient and cost-effective solution for mass production.

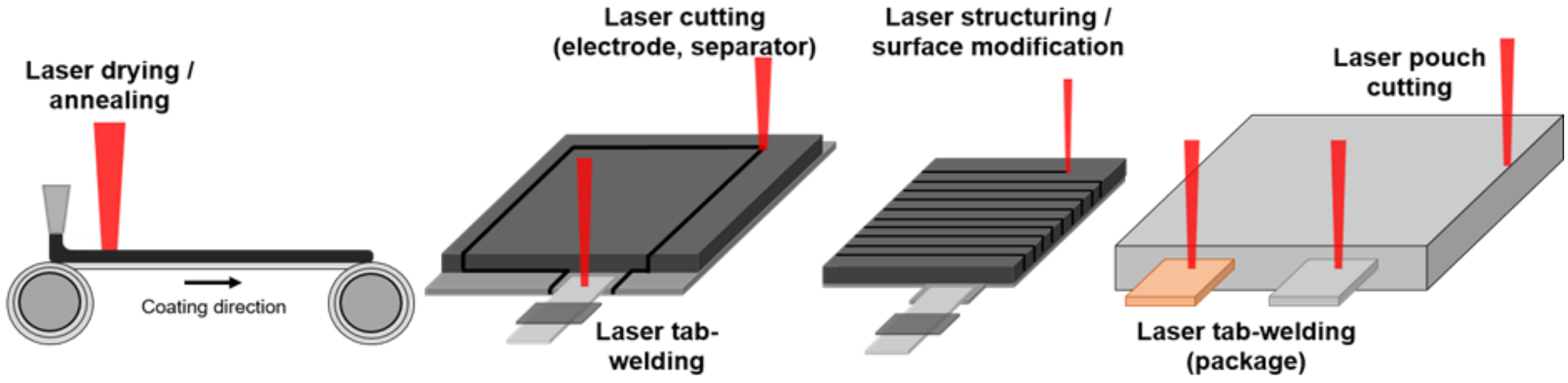

Figure 33.19: Various possible laser applications in LIBs manufacturing. Reprinted from [Zhu, 2024], P. Zhu. Development of three-dimensional lithium-ion cells with water-based Li(Ni0.6 Mn0.2 Co0.2)O2 thick-film cathodes using ultrafast laser ablation. Karlsruher Institut für Technologie (KIT) repository, PhD thesis. Karlsruhe (2024), under Creative Commons Attribution-ShareAlike (CC BY-SA) 4.0 International License (https://doi.org/10.5445/IR/1000175921).

Pulsed nanosecond lasers with wavelengths ranging from 500 to 1070 nm are typically used in laser electrode cutting and the ablation width is generally below 40 µm [Lee, 2020; Berhe, 2022]. Laser-assisted drying of electrode coatings is an alternative to conventional drying with convection air and ovens. However, technique remains at the laboratory scale and requires further development [Neb, 2022].

Laser structuring of current collectors has been achieved through methods like direct laser interference patterning (DLIP) [Zwahr, 2023] and direct laser ablation using lasers with wavelengths of 1030 nm, 515 nm [Zhang, 2017], and 355 nm [Dong, 2021]. These techniques can enhance the adhesion strength between composite graphite anodes and laser-modified Cu current collectors [Zhang, 2017]. However, their electrochemical performance depends on the laser parameters and process conditions. For example, Cu foils structured using a pulsed UV laser with a wavelength of 355 nm and coated with lithium showed reduced internal stress and stabilized SEI during cycling. Full cells incorporating lithium-deposited laser-modified Cu foils and LFP cathodes achieved 45 mAh/g higher capacity after 600 cycles compared to those with unstructured Cu foils [Dong, 2021].

Laser structuring can also be applied to separators. Polypropylene (PP) separators coated with $Al_2O_3$ were processed using femtosecond lasers with a 515 nm wavelength, and different patterns such as micro-channels with 50 µm pitch or through-holes with a 300 µm spacing were generated. Full cells containing NMC 111 cathodes and graphite anodes with laser-structured separators showed varied performance: cells with micro-channel-structured separators retained over 95% capacity after 200 cycles, whereas those with unstructured separators retained 85% capacity under the same conditions. In contrast, separators with through-holes experienced severe capacity drops, retaining 80% capacity after 110 cycles [Pröll, 2015].

However, laser structuring of electrodes is at an early stage of development in frame of the 3D battery concept and has several major advantages in terms of cell performance and battery production, which will be discussed in more detail in the next sections.

# 5.6 3D Battery Concept and Impact on Electrolyte Wettability and Diffusion Kinetics

Long et al. [Long, 2004] introduced the three-dimensional (3D) battery concept for micro-batteries, aimed at designing 3D electrodes that provide shortened lithium-ion diffusion pathways between electrodes, thereby boosting lithium-ion diffusion kinetics. Based on this concept, laser ablation was employed to generate 3D electrodes with various design types. This technique was first introduced for LIBs by the Karlsruhe Institute of Technology (KIT) in 2008 for thin-film batteries [Ketterer, 2008]. Magnetron-sputtered lithium-cobalt-oxide (LCO) and lithium manganese oxide (LMO) thin compact films were structured with UV excimer laser radiation operating with pulse durations of 4 to 6 ns. The laser structuring produced cone-shaped and free-standing microstructures, resulting in an estimated sixfold increase in active surface area [Kohler, 2009].

This classic concept for thin film batteries has been scaled up for practical application by using it to create a new electrode architecture for thick film, i.e., high energy, composite electrodes with the initial primary objective of boosting the electrolyte wetting of porous thick-film electrodes [Pfleging, 2022]. The result is a tandem effect: an increase in electrochemical performance and a significant advantage for battery production in terms of the quality-determining electrolyte filling process, schematically illustrated in Figure 33.20.

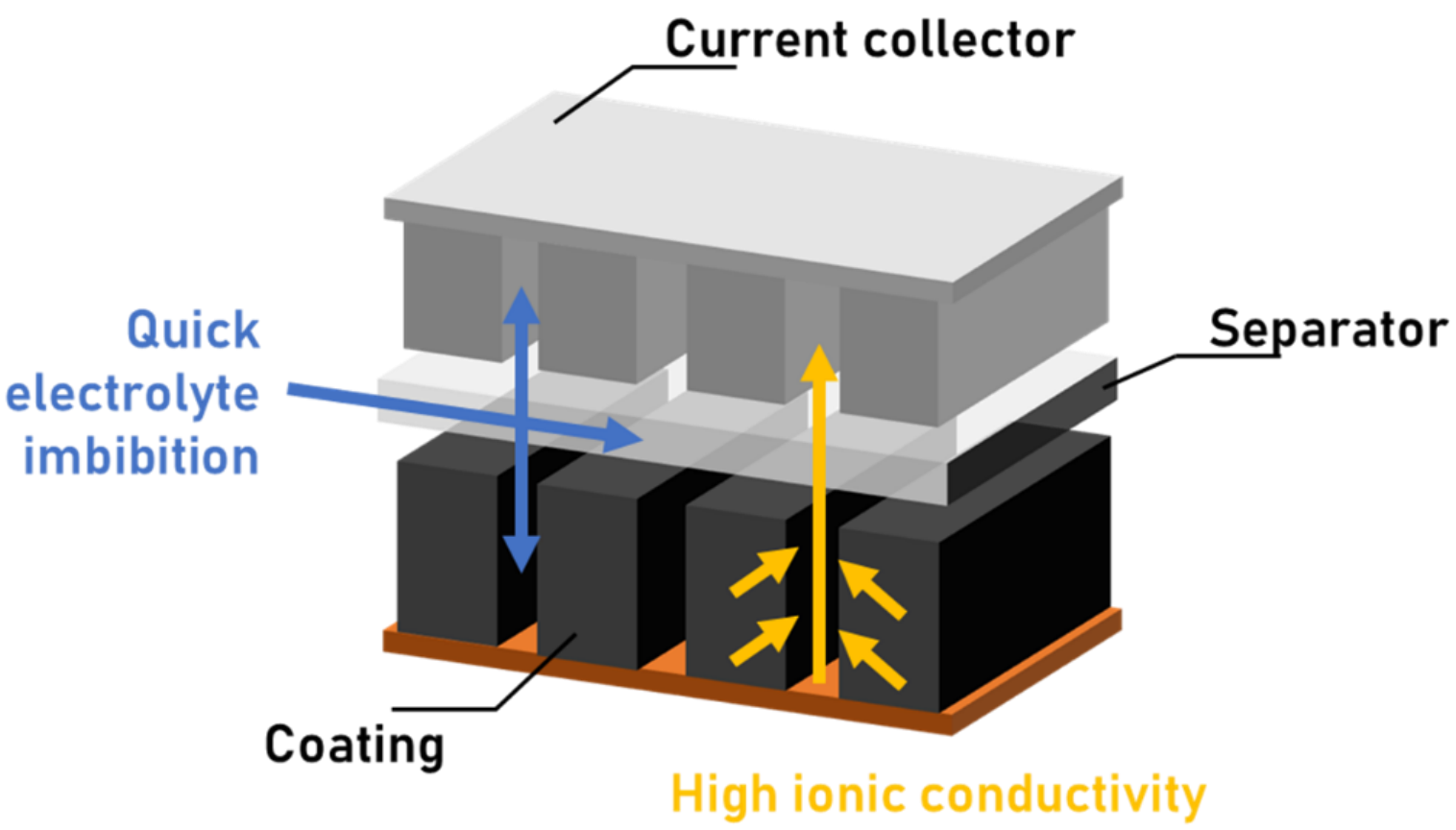

Figure 33.20: Schematic illustration of 3D battery concept with improved electrolyte wetting and ionic conductivity in laser structured electrodes with a line pattern. © KIT

This approach was first realized for NMC 111 and LMO cathodes. Line patterns acting as capillary structures were introduced by laser ablation. Laser-generated micro-grooves in the electrodes act as capillary structures for liquid electrolyte which can be very well described by the classical Washburn equation for capillary rise [Pfleging, 2014]. Nano- and micro-sized pores are instantly filled, enabling rapid and uniform wetting of the entire electrode within seconds, whereas unstructured electrodes exhibit no significant wetting.

Moreover, it was found that using laser-structured electrodes could eliminate the need for prolonged battery storage after assembly, offering the potential to reduce processing costs during electrolyte wetting and subsequent warm aging while lowering the scrap rate. Similar findings were reported using in situ neutron radiography on pouch cells, where cells with laser structured electrodes having grid patterns achieved complete wetting within 15 min, whereas those with unstructured electrodes only reached 49% wetting in the same time frame [Habedank, 2019a].

Further study investigated laser-structured LFP cathodes with line patterns of varying aspect ratios and found that the wettability of liquid electrolyte was enhanced with increasing aspect ratio [Berhe, 2023]. In addition, multiphysics simulations verified that for structured electrodes with line patterns and a pitch of 150 µm, the complete electrolyte wetting takes approximately 14 s, whereas over 950 s is needed for the unstructured electrode to reach saturation of electrolyte. Thus, the electrolyte wetting time is reduced by 98.5% after applying laser structuring [Biebl, 2024].

Tape-cast NMC 111 and LMO electrodes with thickness of 55 µm and 103 µm respectively, were structured with a line pattern having a pitch of 200 µm using a nanosecond Yb:YAG fiber laser [Kohler, 2013]. The laser structuring of LFP composite cathodes was studied and the pulse duration was varied from 350 fs to 200 ns [Mangang, 2016]. It was found that the aspect ratio, defined as the relation of ablation depth and channel width, increased as the pulse duration decreased. A recent study on the structuring of LFP cathodes using ns laser radiation suggested that applying multiple laser ablation passes at low laser fluence is effective in achieving high aspect ratio structures [Park, 2021]. Moreover, through-holes with a diameter of 20 µm were created using laser drilling in composite LFP cathodes and graphite anodes, applying picosecond pulsed laser radiation [Tsuda, 2018]. This technique was extended to double-sided coated LFP cathodes demonstrating enhanced performance under high current conditions [Matsumoto, 2020].

Thick-film NMC 622 cathodes, with thicknesses from 91 to 250 µm, were structured using a femtosecond pulsed laser with line patterns having pitches of 200 to 600 µm. These structured electrodes demonstrated increased energy density compared to unstructured electrodes when discharged at high currents in less than one hour [Zhu, 2019]. NMC 532 cathodes with thicknesses ranging from 100 to 210 µm and electrode porosities of 26% and 50% were laser structured using the same pitch of 200 µm. It was shown that laser structuring, which generates expanded lithium-ion diffusion pathways, can mitigate the reduction in power and energy density typically associated with thick film electrodes [Park, 2019].

Similarly, line-patterned NMC 532 cathodes with ablation depth varying from 50 to 140 µm and different pitches were fabricated, and it was found that groove depth is a more critical parameter than pitch to maximize the electrochemical performance at high current conditions [Park, 2022a]. In addition, the 3D patterning of electrodes with varying thicknesses yielded differing aspect ratios and material losses, the aspect ratio increased four times as the film thickness increased from 40 to 100 µm, while the mass loss decreased by 4% with increasing film thickness [Tran, 2022].

So far, the patterns that appear most often in the literature are line and hole structures, as displayed in Figure 33.21a and b. However, grids, hexagonal structures, or other combined geometries have also been reported for thick film NMC 622 cathodes using a pulsed fiber laser with a pulse duration of 600 fs, as shown in Figure 33.21c and d [Zhu, 2024a].

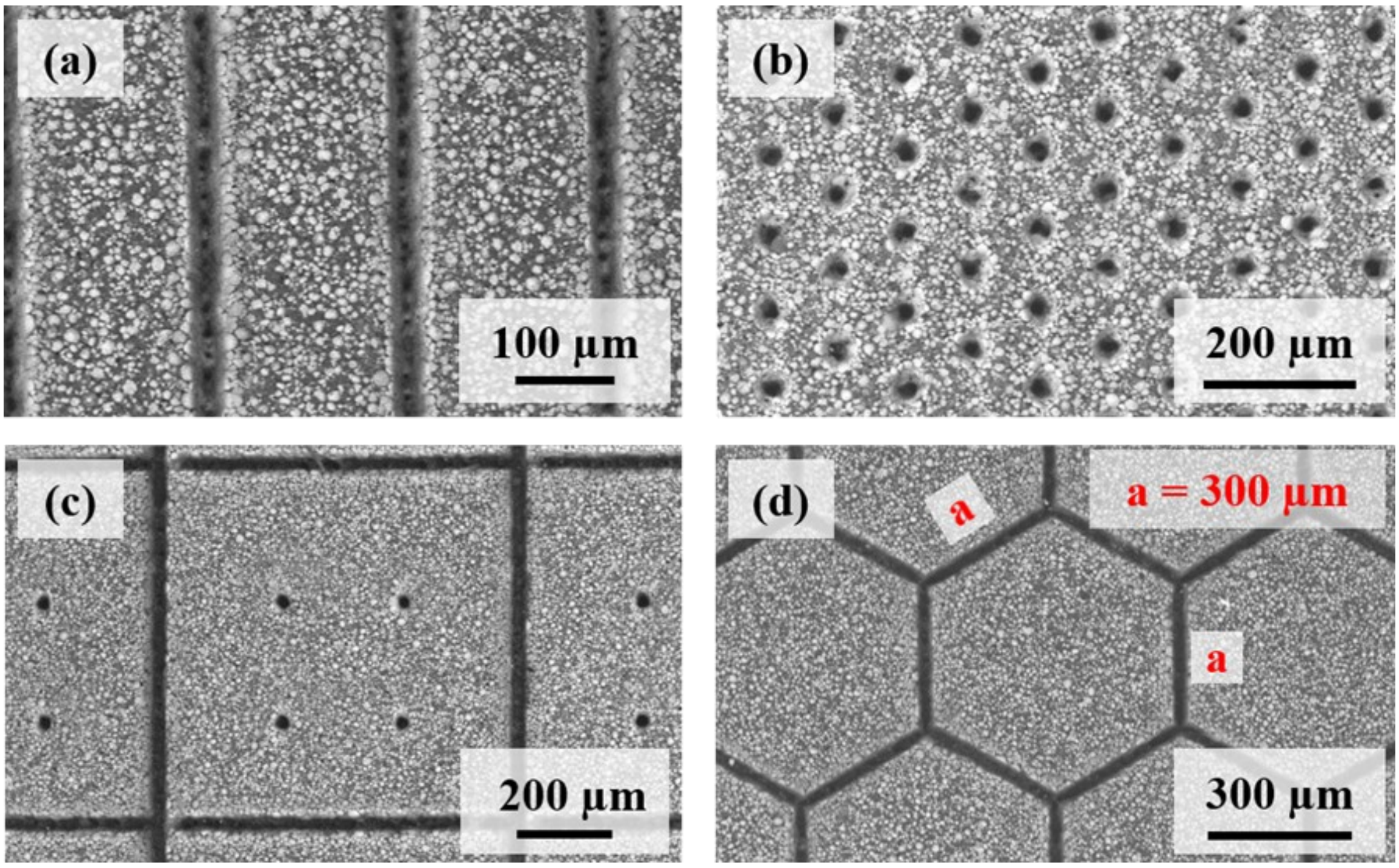

Figure 33.21: Laser structured thick film NMC 622 cathodes with different patterns: (a) lines, (b) holes, (c) grids and holes, (d) hexagonal structures. Reconstructed from [Zhu, 2024a], P. Zhu, B. Ebert, P. Smyrek, and W. Pfleging, The Impact of Structural Pattern Types on the Electrochemical Performance of Ultrathick NMC 622 Electrodes for Lithium-Ion Batteries. Batteries, 10(2), 58 (2024), under Creative Commons Attribution (CC BY) 4.0 License (https://doi.org/10.3390/batteries10020058).

Laser structuring of graphite anodes with a film thickness of 50 μm was first introduced in 2018, using a femtosecond laser with a pulse duration of 400 fs and a wavelength of 1040 nm was employed [Habedank, 2018]. The study showed that high laser peak fluences are more critical for efficient ablation than the total accumulated energy input into the electrode over time. Similar results were reported for the structuring of graphite anodes using a high-power picosecond laser (pulse duration of 12 ps) with a maximum power of 400 W [Meyer, 2022].

Graphite anodes with areal capacity over 4.3 mAh/cm² were structured using a high-power femtosecond laser operated at a wavelength of 1030 nm, while various patterns including lines, grids, and holes were investigated to determine the optimal geometry for graphite anodes [Sterzl, 2024]. Moreover, grid and line patterns were applied for the laser structuring of graphite anodes and blended composite anodes with silicon and graphite (Si/C) [Zheng, 2019]. The laser-generated free spaces shorten lithium-ion pathways, enhancing diffusion kinetics along the interfacial area. Additionally, the mechanical stress resulted from the volume change of Si in Si/C during charging/discharging was significantly reduced after applying 3D structured electrodes.

Most research has employed lasers with pulse durations in femtosecond or picosecond range for electrode structuring due to their minimal heat-affected zone (HAZ). However, some studies also explored lasers with nanosecond pulse duration for this purpose. For example, a pulsed laser with a pulse duration of 0.15 ns was applied to generate hole structures in graphite [Hille, 2023a]. In addition, a nanosecond pulsed laser (15 ns) operating at a wavelength of 532 nm was applied for the structuring of LFP cathodes with hole patterns [Yamada, 2023], while the morphology of groove structures in LFP cathodes generated using a nanosecond pulsed laser (4 ns) at 1064 nm was investigated [Berhe, 2024]. It was concluded that using low laser fluence with multiple scan passes minimizes material loss and can generate grooves with high aspect ratios [Park, 2021].

## 5.7 Impact on Fast Charging and Battery Safety

One main challenge for lithium-ion batteries used for electric vehicles is the charging time. To specify the applied electrical current and the charging or discharging speed of a battery, the term "C-rate" $nC$ [1/h] is introduced and defined as follows:

$$nC = \frac{I_{cell}(t)}{Q_{nom}} \tag{33.2}$$

where $I_{cell}(t)$ [A] is the currently applied current, and $Q_{nom}$ [Ah] the nominal capacity of a cell. Achieving a charging time of less than 15 min to reach approximately 80% SOC, corresponding to a C-rate of 3.2C to 4C, is considered essential to bridge the refueling time gap between electric vehicles and conventional ones [Weiss, 2021]. Hereby the term SOC is defined as the ratio of the currently available capacity $Q(t)$ [Ah] to the maximum stored charge $Q_{max}$ [Ah] in a cell:

$$SOC = \frac{Q(t)}{Q_{max}} \cdot 100\% \tag{33.3}$$

Its value ranges from 0% to 100%, where 0% indicates a fully discharged cell, and 100% indicates a fully charged cell.

Laser-structured graphite anodes with hole patterns provided about 20% higher capacity at 2C to 5C [Habedank, 2018]. By applying fast charging at 4C, 16% higher state of charge (SOC) was reached for cells with hole structured graphite anodes compared to those with unstructured electrodes. Moreover, capacity retentions over 91% and 86% after 600 cycles were observed for cells with hole patterned graphite anodes under 4C and 6C fast-charging conditions, respectively [Chen, 2020].

Laser structuring of graphite anodes using hole patterns significantly reduced cell impedance at different temperatures ranging from -15 °C to 25 °C, particularly the charge transfer resistance, and is feasible to minimize or prevent lithium plating. This means laser structuring of anodes enables faster charging of electric vehicles at low temperatures without the risk of inducing safety issues caused by lithium plating [Habedank, 2019b].

Graphite anodes with a thickness of 75 µm were laser structured using a line pattern with a 100 µm spacing. The cells showed excellent rate capability up to 3C [Zheng, 2019]. Ultrathick graphite anodes with an areal capacity of 20 $mAh/cm^2$ (film thickness of 410 µm) were structured using line patterns, and a 90% reduction in ionic resistance and enhanced rate capability at C/20 to C/5 were observed after laser structuring, which was attributed to the carbonization of the PVDF binder [Park, 2022b].

A recent study investigated different pattern geometries, such as line, grid, and hole, on graphite anodes and found enhanced rate capability at C-rates ranging from 1C to 3C and lower ionic resistivity compared to cells with unstructured electrodes [Sterzl, 2024]. In addition, the lithium plating onset was shifted towards higher C-rates after employing 3D electrodes. For example, cells with line or hole structured graphite anodes started to show lithium plating at 3C and 2C, respectively, while those with unstructured electrodes began to have lithium plating at 1C.

Moreover, cells containing hole structured graphite anodes showed lower capacity at high C-rates (2C to 5C) compared to cells with line structured electrodes. This suggest that the wetting and rewetting of the electrode with electrolyte during cycling likely affect the lithium plating behavior, since there are no capillary structures in the hole structured electrodes.

The success of 3D graphite electrodes can be extended to Si/C composite anodes, where cells with grid-structured Si/C anodes demonstrated higher capacities at C/5 to 2C during rate capability analysis and over 500 mAh/g more capacity after 100 cycles at C/5 [Zheng, 2019]. Post-mortem analysis of morphology and lithium concentration using laser-induced breakdown spectroscopy (LIBS) analysis revealed that the mechanical stress in Si/C composite anodes can accelerate chemical degradation, leading to rapid cell failure, short battery life and spontaneous cell failure.

In other research work, it was shown that by implementing 3D architectures in Si/C electrodes using laser ablation, rate capability and lifetime are enhanced, and the electrode showed a homogenous lithium distribution after charging at 1C (see Figure 33.22) [Zheng, 2022]. The 3D lithium distribution in Figure 33.22 clearly illustrates the effect of diffusion kinetics for the fast charging of 2D and 3D LIBs and in particular for batteries with high-energy Si/C anodes. During lithiation of the unstructured Si/C electrode, only the upper layers could store a significant amount of lithium, while the lower layers close to the current collector provided almost no capacity to the overall cell due to diffusion kinetic limitations imposed by the large volume expansion. For the structured electrode, however, a homogeneous lithium distribution was achieved over the entire electrode. This demonstrates the avoidance of a diffusion overpotential in 3D batteries and, in particular for the Si/C electrode, the successful suppression of mechanical degradation that would normally be expected in conventional batteries due to the high-volume changes associated with the alloying of Si with lithium.

## 5.8 Impact on High-Power Operation and Lifetime

The power density of cells can be increased by structuring the cathodes. Galvanostatic analyses revealed that half cells with structured NMC 622 cathodes showed higher specific capacity at discharge rates of C/2 to 3C [Zhu, 2019]. For example, electrodes with thicknesses of 151 µm and 250 µm, structured with a 200 µm line pitch, showed a 40 mAh/g higher discharge capacity at C/2 and 1C, respectively. Additionally, cells with 250 µm thick NMC 622 electrodes achieved higher capacities than those with unstructured electrodes at C-rates above C/5. As the electrode thickness increased, the optimal C-rate for structured electrodes shifted to lower values.

In another study, thick film NMC 622 cathodes with a thickness of 150 µm were processed with water-soluble binders and subsequently structured using laser ablation [Zhu, 2023]. It was shown that cells with phosphoric acid modified laser structured NMC 622 cathodes exhibited 10 to 30 mAh/g higher specific capacity at C/2 and 1C compared to the reference cells containing NMC 622 cathodes with conventional PVDF-based electrodes. This concept was further developed and verified in pouch cells with NMC 622 cathodes and graphite anodes, where the energy density at power densities ranging from 140 to 1300 W/L increased following aqueous slurry processing and laser structuring on both, cathodes and anodes [Zhu, 2024b]. Moreover, combining laser structuring with cathode slurry pH modification extended cell lifetime by a factor of four, compared to the reference cells with 2D electrodes using PVDF binder.

A recent study compared eight different laser structuring patterns for NMC 622 cathodes while maintaining the same mass loss [Zhu, 2024a]. The results showed enhanced rate capability at C-rates above C/2 for all structured NMC 622 cathodes compared to those with unstructured electrodes, regardless of the pattern. However, cells with hole-structured cathodes had the lowest capacity increase at 1C than those with other patterns, while cells with hexagonal structured electrodes achieved the maximum capacity gain at 2C. Furthermore, line-like patterns with a 200 µm pitch were applied for the structuring of NMC 811 cathodes with layer thicknesses between 40 µm and 100 µm, and cells with thick-film cathodes exhibited lower charge transfer resistance and delivered three times higher capacity at 2C compared to those with unstructured electrodes [Tran, 2022]. LIBS analyses revealed that laser-generated channels create additional lithium-ion diffusion pathways from the liquid electrolyte into the active materials. As a result, the lithium concentration in cycled 3D thick film electrodes becomes more homogeneous compared to unstructured electrodes, especially at high C-rates [Smyrek, 2019].

Pouch cells with laser structured NMC 111 cathodes with a film thickness of 55 µm and line patterns reached 80% SOC after 2290 cycles, which is 16 times longer than the cycle life of cells with unstructured electrodes [Pfleging, 2014]. As for pouch cells containing laser structured graphite anodes with hole patterns and unstructured NMC 111 cathodes, over 170 additional cycles were achieved before reaching their 80% SOC than those with unstructured electrodes [Kriegler, 2021]. Furthermore, an increase of up to 51% in cycle lifetime was reported for pouch cells containing laser structured Si/C composite anodes with line patterns and unstructured NMC 622 cathodes [Meyer, 2021].

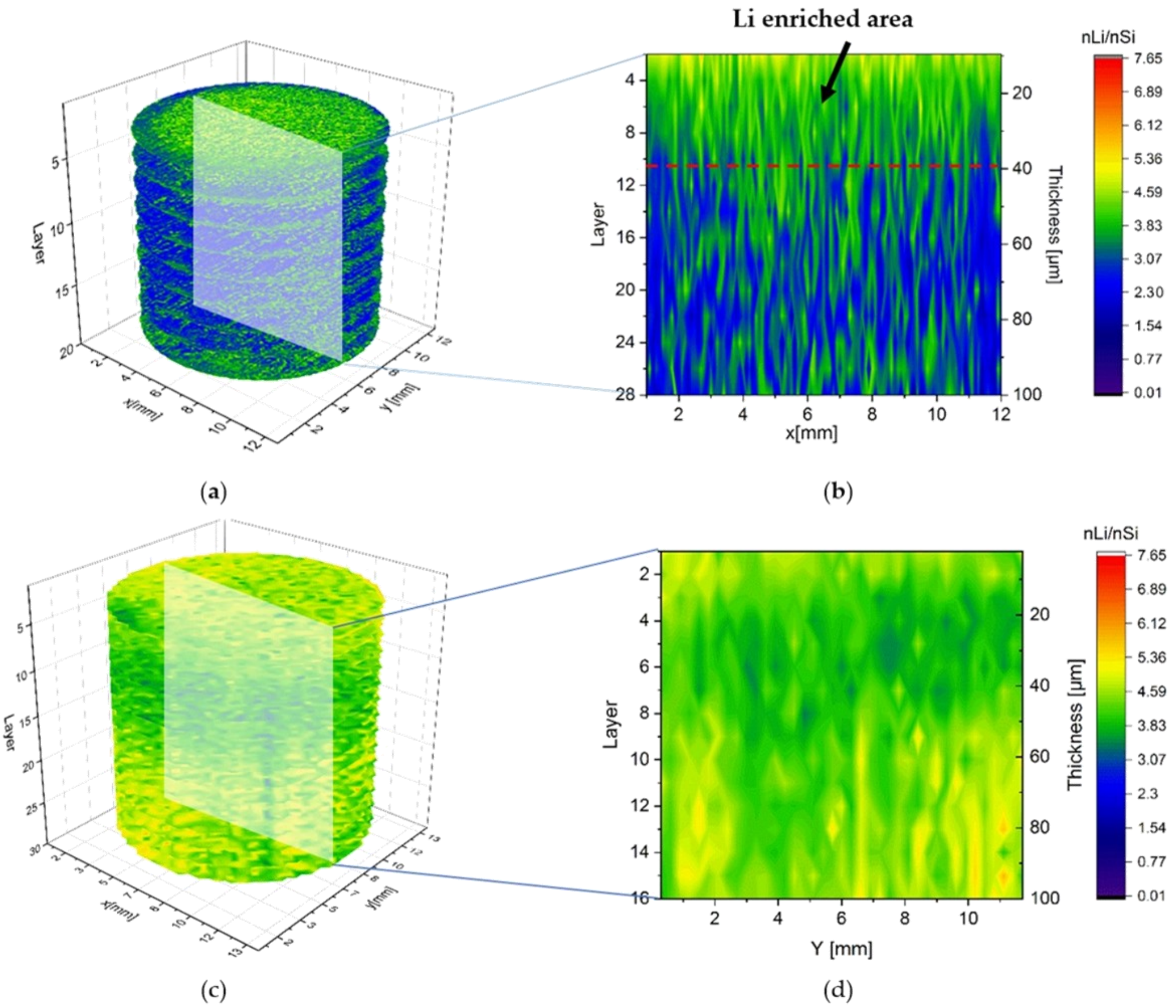

Figure 33.22: Lithium concentration profile in structured and unstructured Si/C electrodes which were delithiated by a C-rate of 1C: (a) 3D illustration of lithium concentration in reference coin cell electrode and (b) its corresponding lithium distribution in cross-sectional view; (c) 3D illustration of lithium concentration in a structured coin cell electrode and (d) its corresponding lithium distribution in cross-sectional view. Reproduced from [Zheng, 2022], Y. Zheng, D. Yin, H.J. Seifert, and W. Pfleging, Investigation of Fast-Charging and Degradation Processes in 3D Silicon–Graphite Anodes. Nanomaterials, 12(1), 19 (2022), under Creative Commons Attribution (CC BY) 4.0 License (https://doi.org/10.3390/nano12010140).

The impact of laser patterning configurations on cathodes and anodes was studied by comparing various combinations of structured/unstructured electrodes with hole patterns. Results showed that cells with both structured anodes and cathodes delivered the highest rate capability from 2C to 5C, while cells with only structured anodes exhibited the second-highest discharge capacity [Hille, 2021]. Similarly, research on coin cells with laser structured NMC 622 cathodes and graphite anodes with line patterns demonstrated that the greatest capacity increase can be achieved when both electrodes are laser patterned [Dunlap, 2022].

# 5.9 High-Energy Cells — Upscaling on Footprint

Laser structuring of electrodes has demonstrated significant advantages in improving energy density, enabling high power applications, fast charging, lifetime, and safety of LIBs and is thus attracting increasing attention. However, to bring this technology to the next technology readiness level (TRL), it is essential to prove this new concept on a larger battery footprint to bridge the gap between laboratory and product-relevant level and to provide application relevant electrochemical data.

Figure 33.23 shows the lifetime analyses of 1 Ah-pouch cells assembled by KIT using NMC 111 cathodes with an areal capacity of around 2.1 mAh/cm² (per side) and graphite anodes (with and without laser structuring). Each cell consisted of four double-side coated cathodes, three double-side coated anodes, and two single-side coated anodes, all with a footprint of around 8 ×8 cm². It is worth mentioning that these battery cells consist of electrodes with a relative low film thickness, where diffusion kinetic limitations are not a bottleneck. It is therefore not surprising that for a low number of cycles, all cells performed similarly. After more than 400 cycles at C/2, the cell with unstructured electrodes retained 74.7% of its initial capacity, while the cell with structured anodes and unstructured cathodes achieved a capacity retention of 86.7%. Notably, the cell having the laser-structured electrodes showed the highest capacity retention of 90.2% by the end of the analysis.

Graphite anodes with an areal capacity of 3.2 mAh/cm² were laser structured with hole patterns and assembled with unstructured NMC 532 cathodes in 2 Ah pouch cells, and fast charging at 6C and 4C was demonstrated [Chen, 2020]. The laser structured graphite anodes with hole-like patterns were further upscaled and paired with unstructured NMC 111 cathodes in 3 Ah pouch cells [Kriegler, 2021]. Moreover, hole structured graphite anodes with an areal capacity of 3.1 mAh/cm² were assembled versus NMC 811 cathodes in 21700-type cylindrical cells, achieving a total capacity of 3.4 Ah. However, the processing time for each double-side coated anode (with an area of about 660 cm²) was approximately 60 minutes [Scheck, 2024]. A recent study confirmed the scalability of the process by employing NMC 622 cathodes (areal capacity of 4.1 mAh/cm²) and graphite anodes (areal capacity of 4.4 mAh/cm²) with line patterns in 15 – 20 Ah industrial-level pouch cells containing nine cathodes and ten anodes [Meyer, 2023]. The results validated the enhancement in rate capability and a threefold increase in cell lifetime of cells with laser structured electrodes compared to those with unstructured electrodes.

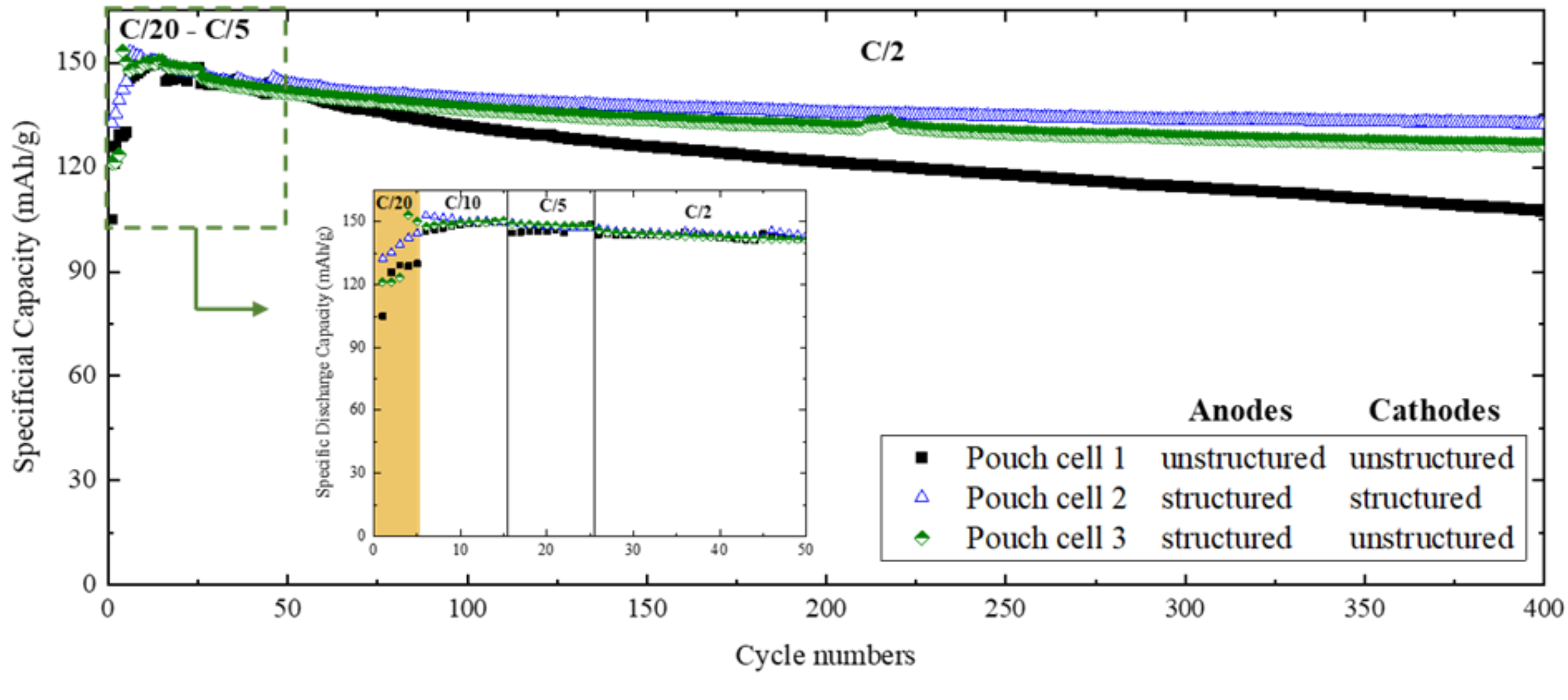

Figure 33.23: The lifetime analyses of 1 Ah-pouch cells with structured/unstructured NMC 111 cathodes and graphite anodes. © KIT.

## 5.10 Laser Process Upscaling

Now that the 3D battery concepts have been proven, the question remains as to how and to what extent the laser process can be scaled up, or if there is another technical approach to enable efficient and rapid 3D structuring of electrodes. The structuring process needs to be adapted to the battery production line. However, it is important to note that the production lines themselves may differ in terms of coating technology, drying methods, mass loading, footprint, and cell format, which in turn will result in individual processing speeds and therefore process requirements for a suitable structuring process.

It is therefore most reliable to first consider a structuring process that provides a stand-alone solution within the battery production environment (≥TRL7). This section focuses on laser structuring of electrodes; however, it is worth noting that several alternative processes, such as printing, mechanical embossing, or templating techniques, which have also been studied and evaluated on lab-scale [Pfleging, 2018, Rist, 2023, Sandherr, 2023, Ezeigwe, 2025].

The last decade has seen significant advancements in high-power ultrashort pulse laser sources. This, in turn, is an essential prerequisite for upscaling the 3D battery concept based on laser ablation. For example, ongoing research and technological development, including new optical concepts for multi-beam processing, has currently enabled laser electrode structuring to advance from laboratory level to a TRL of 5, approaching a relevant status for industrial feasibility and applicability. A further increase to TRL 7 would consolidate its technological potential and pave the way for its integration into battery production with Ah-scale cells. Recent studies have explored various methods to scale up laser structuring. Increasing laser power (or laser fluence) leads to higher ablation rate. Tthus, the total processing time can be reduced. This has been observed for the structuring of LFP cathodes with line patterns [Park, 2021] and large areal laser structuring of graphite anodes with line patterns [Sterzl, 2023].

The laser scan speed $v_s$ [mm/s] related to average laser power $P_L$ [W] can be described as [Gafner, 2021]:

$$v_s = \frac{4}{\pi e^2} \cdot \frac{1-PO}{(d_f/2)\Phi_{th}} \cdot P_L \cdot 100 \tag{33.4}$$

where $PO$ [-] is the pulse overlap, $d_f$ [mm] the laser spot diameter, $\Phi_{th}$ [J/cm$^2$] the threshold laser fluence for ablation. Thus, increasing the average laser power results in an increase of the scan speed for the same laser spot size and constant pulse to pulse overlap. However, the ablation width increases simultaneously with increasing average laser power, resulting in a reduced aspect ratio [Sterzl, 2023].

Another approach to accelerate the laser scan speed is to apply a polygon scanner system [Habedank, 2020]. This setup allows scan speeds up to 800 to 1000 m/s, which is up to fifty times higher than the scan speed of conventional galvanometer scanners. Employing optical elements to split the laser beam into multiple beams is also feasible to increase the laser process efficiency. For instance, by combining a multibeam approach with a kilowatt-level high-power laser, the processing time for laser structuring of electrodes has been estimated to be significantly reduced [Pfleging, 2020; Sterzl 2023; Hille, 2023b]. In addition, laser structuring with beam shaping element is another option and showed higher ablation efficiency compared to the single-beam method [Geiger, 2024]. Another innovative laser drilling technique for electrodes involves using a hollow cylinder with openings and a laser source positioned inside the cylinder. This method has been reported to generate up to 100,000 holes per second in the electrodes and is compatible with roll-to-roll (R2R) processing [Yamada, 2023]. Another strategy to reduce processing time is to increase the laser repetition rate. However, as the repetition rate increases, a reduction in ablation efficiency per laser pulse must be considered as the laser intensity for ablation may be attenuated by the material vapor plasma generated [Habedank, 2018; Meyer, 2022; Sterzl, 2023].

Pulse burst processing can lead to enhanced ablation efficiency, as shown for MHz burst mode for LFP and NMC 111 cathodes. Hereby, the ablation rate was 2.5 times and 5 times higher, respectively compared to a single pulse of the same energy [Trenn, 2024]. The ablation per burst and ablation efficiency increased with longer pulse bursts. Similar results for MHz bursts have been reported for graphite anodes and NMC 811 cathodes, where the processing time was reduced by a factor of 4 [Sikora, 2024]. More recently, promising results have been obtained using high power (440 W) ultrashort pulse (USP) lasers with GHz burst repetition rates for electrode structuring with a significant increase in laser ablation efficiency in comparison to single pulse laser ablation [Sterzl, 2025]. In addition to ablation efficiency, the generated topography, i.e., aspect ratio and resulted mass loss, plays a key role for practical use in LIBs, which in turn requires precise adjustment of the laser parameters based on the process conditions and material properties.

# 6 Conclusion and Outlook

As demonstrated in this chapter, laser technology has proven to be a powerful and versatile tool in advancing energy generation, storage, and conversion. The following conclusions can be drawn, highlighting the key benefits, challenges, and future potential of laser-based processes in driving sustainable and efficient energy applications. The recent rapid developments in the technology and upscaling of the average power of ultrashort pulsed lasers into the kW regime (see Chap. 2 (Gailevicius et al.)) enabled a boost in such applications and will further drive laser-based green energy solutions.

**Laser Fusion for Energy Generation:** Laser-driven fusion is at the cutting edge of the quest for limitless, clean energy. By using high-power lasers to achieve inertial confinement fusion (ICF), scientists are currently working to recreate the conditions of stellar fusion on Earth for providing and controlling a powerful and sustainable energy supply and logistics, while avoiding the nuclear waste problem of the conventional nuclear technology. Milestones like the National Ignition Facility’s breakthrough in achieving scientific breakeven in 2022 highlight the potential of this approach, though challenges remain in improving energy efficiency and controlling plasma instabilities. The precision tuning of laser parameters, such as intensity, pulse duration, wavelength, and beam uniformity, has been pivotal in enhancing fusion yields. Advances in direct- and indirect-drive methods, along with sophisticated laser configurations, continue to push the limits of what is possible. Global research facilities, including NIF, LMJ, and Omega, serve as key hubs for advancing high-energy-density physics and refining fusion techniques. Despite the obstacles, laser-driven fusion remains one of the most promising but also highly challenging pathways to sustainable energy. With ongoing technological advancements particularly in laser technology and materials sciences, it has the potential to revolutionize power generation, bringing humanity closer than ever to harnessing the energy of the stars.

**Laser Technology in Solar Energy Harvesting:** Laser technologies will continue to revolutionize the photovoltaic industry, with significant advances expected in precision, efficiency, reduction of manufacturing costs and scalability. The laser-based manufacturing technologies discussed in this chapter, although not providing an exhaustive overview of all laser-based methods utilized in photovoltaics, represent a balanced selection of established industrial techniques (e.g., laser scribing) and emerging, cutting-edge approaches for next-generation photovoltaic applications (such as LIFT for micro-concentrator photovoltaics). It can be expected that in the near future several innovations will further be developed and integrated into industrial manufacturing. For instance, the deployment of femtosecond laser sources for scribing thin film modules might further improve the quality of the scribes enabling thinner and more efficient and reliable devices. Furthermore, fs-laser pulses can widely avoid heat-induced damage in more sensitive films such as perovskite absorbers. It has been shown that efficiency improvements through optimized light management via laser texturing can be achieved not only in wafer-based silicon solar cells but also in thin-film devices. However, for the successful transfer of these approaches to the industry, important challenges should be overcome. On the one hand, efficient post-processing surface passivation is vital to avoid electronic defects that could diminish the efficiency gains achieved through enhanced light trapping. On the other hand, surface texturing throughputs reported in the literature still do not meet industry standards, and therefore high-speed scanning systems and processes should be further developed to bridge this gap and enable large-scale deployment. Bottom-up laser-assisted manufacturing of micro-concentrator solar cells may additionally help to reduce material consumption, while enhancing solar cell efficiencies. It is important to highlight that in the coming decades, a huge volume of silicon-based solar modules will reach the end of their operational lifespan, thereby driving the urgent need for the development of efficient and sustainable recycling strategies. Here, laser processing represents a promising approach for the selective debonding and separation of valuable metals and polymers with high precision and minimal environmental impact. Consequently, laser-assisted recycling methods are anticipated to assume an increasingly vital role in addressing this pressing challenge and contributing to a more circular photovoltaic economy.

**Laser Techniques for Improved Electrochemical Performance:** We have shown that laser-based microprocessing techniques can be of relevance for various electrochemical applications, particularly in power-to-gas technologies. These methods enable not only the geometrical enlargement of electrode surface areas but also the precise tailoring of surface chemistry to achieve distinct wetting characteristics with liquids. Such modifications play a crucial role in enhancing overall electrochemical performance, making laser processing a promising approach for optimizing energy conversion processes. Despite these advantages, several critical challenges must be addressed before laser-based surface structuring can be successfully transferred to industrial applications. One of the primary concerns is the long-term stability and performance of laser-structured electrodes. The additional capital expenditures (CAPEX) and associated costs related to laser processing currently present a substantial economic barrier. To justify these expenses, significant improvements in process efficiency and durability must be demonstrated. Moreover, since the relative long processing time directly contributes to the elevated costs of laser structuring, reducing it is essential for improving economic feasibility. Advancements in laser technology, such as higher processing speeds and more efficient beam delivery systems, will be crucial in making these techniques more competitive for large-scale applications. Although advancements in polygon scanner technology have led to considerable improvements in throughput, the primary bottleneck remains the laser source itself. To meet industrial demands, laser systems must operate at extremely high repetition rates in the megahertz (MHz) range while maintaining pulse energies exceeding 1 mJ. Overcoming these technological limitations will be essential for the successful industrial implementation of laser surface texturing. If these challenges are resolved, laser-based structuring techniques could make a significant contribution to the ongoing energy transition by enhancing the efficiency and durability of electrochemical systems, ultimately supporting the broader adoption of sustainable energy solutions in modern society.

**Laser Technology for Advanced Battery Systems:** One of the medium- to long-term developments in battery production will be an increase in the active mass per electrode sheet (thick film concept). This allows a significant increase in energy density of 20 to 30% at the cell level [Pfleging, 2020]. This development is flanked by the further development of active materials such as Si/C or lithium-metal anodes, and on the cathode side, high-nickel NMC and high-voltage materials such as LiNi0.5Mn1.5O4 (LNMO) [Reinhold, 2024]. However, the thick film concept has a significant disadvantage with respect to lithium-ion diffusion kinetics, which can only be counteracted by appropriate measures in electrode structuring and cell design development. Ultrashort-pulsed high power laser structuring has proven to be suitable to achieve excellent high-rate performance even in high energy cells. First concepts for upscaling with multi-beam concepts and roll-to-roll process control have been successfully implemented. The further increase of the TRL and the transfer to pilot plants are the subject of current research and development. Optics, scanners, optical concepts for beam guidance, and sensor concepts for quality assurance must be further developed, as well as the suitable design of high-power laser beam sources for the anode- and cathode-specific process windows. At the same time, returning the active material removed by the structuring process or upon recycling into the value chain is another important development step to produce new batteries. This can further reduce costs and will realize a closed loop in the sense of enhanced sustainability. The transferability of the 3D battery concept for next generation batteries and solid-state batteries has already been tested in preliminary studies and will strengthen the research field in the long term.